\documentclass[prd,amsmath,floats,amssymb,authoryear,floatfix,superscriptaddress,nofootinbib,onecolumn]{revtex4} 
\usepackage{graphicx}
\usepackage{hhline} 
\usepackage{caption}
\usepackage{natbib}
\usepackage{appendix}
\usepackage{multirow}
\usepackage{makecell}
\usepackage{placeins}

\renewcommand{\thetable}{\textbf{\arabic{table}}}

\makeatletter
\renewcommand{\fnum@figure}{\textbf{Figure~\thefigure}}
\renewcommand{\fnum@table}{\textbf{Table~\thetable}}
\makeatother

\newcommand{\rthis}[1]{\textcolor{black}{#1}}
\usepackage{float}
\usepackage[caption=false]{subfig}
\usepackage[paperwidth=210mm,paperheight=297mm,centering,hmargin=1cm,vmargin=2.5cm]{geometry}

\usepackage[plainpages=false, colorlinks=true, anchorcolor=blue, linkcolor=blue, citecolor=blue, bookmarks=false]{hyperref}
\begin{document}
\preprint{APS/123-QED}

\title{Cosmological Constraints on Neutrino Masses in a Second-Order CPL Dark Energy Model}

\author{Shubham Barua}
 \altaffiliation{Email:ph24resch01006@iith.ac.in}
\author{Shantanu Desai}
 \altaffiliation{Email:shntn05@gmail.com}
\affiliation{
 Department of Physics, IIT Hyderabad Kandi, Telangana 502284,  India}




\begin{abstract}
Recent DESI results indicate a strong preference for dynamical dark energy (DE) when baryon acoustic oscillation (BAO) measurements are combined with supernovae (SNe) and cosmic microwave background (CMB) data using the Chevallier-Polarski-Linder (CPL) parameterization. We analyze the exponential (EXP) parameterization, which introduces a second-order correction to CPL. We determine and compare the 95\% upper bounds on the sum of neutrino masses for three dark energy (DE) models---$\Lambda$CDM, CPL, and EXP---across four neutrino mass hierarchies (1 massive/2 massless, degenerate, normal, inverted) and multiple dataset combinations (CMB$+$BAO, CMB$+$BAO$+$PantheonPlus, CMB$+$BAO$+$DESY5), employing both Bayesian and frequentist frameworks with physical lower limits from oscillation experiments (0.059 eV and 0.11 eV). Our results show that CPL yields tighter ($\lesssim 10$\%) bounds compared to EXP. We further confirm earlier findings that neutrino mass constraints are only mildly sensitive to the assumed hierarchy and that the frequentist bounds are tighter than Bayesian ones. Furthermore, the  imposed oscillation lower limits, the datasets used and the DE parameterizations play a crucial role in the inferred cosmological neutrino mass bounds. For the datasets, hierarchies, and DE parameterizations considered, we find no statistically significant evidence for nonzero neutrino mass consistent with oscillation lower limits.
\end{abstract}


\maketitle
\section{Introduction}
\label{sec:1}

Despite being fundamental components of the Standard Model of Particle Physics, neutrinos remain among its most mysterious particles, with key properties such as their absolute masses and true nature still unknown~\cite{gershtein_1966, Lyubimov_1980,PDG}. Their contribution to cosmology, especially the evolution of the universe, is non-negligible~\cite{ cowsik_1973, lesgourges_2006, loverde_2024, lesgourgues_2013, lesgourges_2012, wong_2011, archidi_2017, vagnozzi_2019}. Generally, the amount of radiation energy density ($\rho_r$) is parameterized by~\cite{abazajian_2015, shvartsman_1969, steigman_1977}:
\begin{equation}
    \rho_r = \left(1+\frac{7}{8}\left(\frac{4}{11}\right)^{4/3}N_\text{eff}\right)\rho_{\gamma},
\end{equation}
where $\rho_{\gamma}$ represents the photon energy density and $N_\text{eff}$ is the contribution of all relativistic species excluding photons. In the simplest case, $N_\text{eff} = 3$~\cite{dodelson_2003, kolb_1990}. This corresponds to the case when neutrinos decouple from the other species instantaneously. Non-instantaneous decoupling of neutrinos and finite temperature QED effects lead to $N_\text{eff} = 3.046$~\cite{hannestad_1995, dolgov_1997, esposito_2000, hannestad_2002, mangano_2002, froustey_2020, bennett_2021, akita_2020, komatsu_2011}. However, determining the individual masses of the three neutrino species remains a challenge. It should be noted that current cosmological observables are primarily sensitive to the sum of the three masses, $M_\text{tot}$~\cite{abajajian_2011, dvorkin_2019, gerbino_2017, lesgourges_2006}. If the three active neutrinos have different masses, then they will each become non-relativistic at different temperatures, and 
will cause different levels of suppression to the small scale matter power spectra. Therefore, the same total mass, but different mass splittings should result in slightly different matter power spectra. However, current experiments do not have the sensitivity to distinguish these differences with reasonable significance~\cite{archid_2020, xu_2021}. Lower limits on the sum of neutrino masses $M_\text{tot}$ have been inferred from terrestrial neutrino oscillation experiments. These experiments measure the mass-squared differences between the neutrino mass eigenstates.  Results from these experiments show that at least two of the neutrino species are massive~\cite{fukuda_1998, ahmad_2002}. The inability to determine the individual neutrino masses ($m_1, m_2$, and $m_3$)~\cite{capozzi_2021, esteban_2020, de_salas_2021} lead to multiple possibilities for the  hierarchy among the masses: the normal hierarchy (NH) [$m_1<m_2<m_3$] and the inverted hierarchy (IH) [$m_3<m_1<m_2$]. In each of these hierarchies, $M_\text{tot} = m_1+m_2+m_3$ can be calculated as follows: 
\begin{align*}
    M_\text{tot} &= m_0+\sqrt{\Delta m^2_{21}+m^2_0} + \sqrt{\Delta m^2_{31}+m^2_0}, \hspace{1cm} \text{(NH)} \\
    M_\text{tot} &= m_0+\sqrt{|\Delta m^2_{32}|+m^2_0} + \sqrt{|\Delta m^2_{32}|-\Delta m^2_{21}+m^2_0}, \hspace{1cm} \text{(IH)}
\end{align*}
where $m_0 = m_1$ for NH and $m_0=m_3$ for IH. Neutrino oscillation experiments measure two mass squared difference values: the atmospheric ones lead to $|\Delta m^2_{3\ell}| = |m_3^2 - m^2_{\ell}| = (2.45\pm 0.04) \times 10^{-3} \text{eV}^2$ ($\ell$ = 1 for NH and $\ell$ = 2 for IH) while the solar ones give $\Delta m^2_{21} = m^2_2-m^2_1 = (7.5^{+0.19}_{-0.17})\times10^{-5} \text{eV}^2$~\cite{esteban_2017, gonzalez_2021}. Using $m_0 = 0$, we can determine the lower bounds on $M_\text{tot}$ in the two hierarchies:
\begin{align*}
    M_\text{tot} & > 0.059\ \text{eV} \hspace{0.4 cm} (\text{NH})\\
    M_\text{tot} & > 0.1\ \text{eV} \hspace{0.4 cm} (\text{IH}).
\end{align*}
However, none of the current neutrino oscillation data seem to find a strong preference for either hierarchies. 

The Dark Energy Spectroscopic Instrument (DESI) collaboration has released Data Release 1 (DR1)~\cite{desi_2024} and Data Release 2 (DR2)~\cite{desi_2025}. Both data releases provide strong evidence for Dynamical Dark Energy (DDE),  when combined with cosmic microwave background (CMB) and Type Ia Supernovae (SNe Ia) data at significance of (2.8 - 4.2)$\sigma$. A large number of works have investigated extended Dark Energy (DE) parameterizations using different dataset combinations and methodologies to determine the preference for DDE~\cite{ling_2025,yashiki_2025, escamilla_2024,wang_2025, dinda_2025, li_2025, huang_2025, tian_2025, yunhe_2025, jun_xian_2025, li_2024, wu_2025, pan_2025, goswami_2025, silva_2025, shubham_2025, colgain1_2025, colgain2_2025, colgain3_2025, mukherjee_2024, mukherjee_2025, velazquez_2024, park_2024, park_2025, ratra_2024, giare_2024, ioannis_2025, gialamas_2025, di_2025, braglia_2025}. DESI also reports tight upper bounds\footnote{In this paper we quote 95\% confidence limits on the neutrino mass sum following literature convention, unless otherwise specified} on the sum of neutrino masses (CMB$+$BAO) $M_\text{tot} \lesssim 0.064$ eV in the $\Lambda$CDM model (assuming three degenerate neutrino mass states)~\cite{elbers_2025}. This bound lies very close to the lower limit inferred from oscillation data (for NH) and is in tension with IH. The KATRIN experiment~\cite{katrin_2025} directly measures the upper bounds on the neutrino mass sum and provides a less stringent upper bound of $M_\text{tot} \lesssim 0.45$ eV ($90\%$ confidence limit). Negative neutrino masses have also been reported by several works which analyzed cosmological data~\cite{naredo_2024, green_2025, elbers_willem_2025, vagnozzi_2018}. 
However, other dataset combinations and extended DE parameterizations can relax these bounds, thereby restoring consistency with oscillation data. 
This happens due to the degeneracy between $M_\text{tot}$ and the equation of state (EOS) of DE~\cite{hannestad_2005, shouvik_2018, zhang_2017, zhang_2016, shouvik_2025, yang_2017, gabriel_2025}. 

In addition to NH and IH, we also consider two widely used neutrino mass approximations: degenerate mass hierarchy (DH) and 1 massive/2 massless neutrinos (1M). In the DH scenario, the three neutrino mass eigenstates are degenerate ($m_1=m_2=m_3=M_\text{tot}/3$)~\cite{hannestad_2016, choudhury_2020, lesgourges_2004}. In the 1M approximation, $m_1=M_\text{tot}, m_2=m_3=0$~\cite{hannestad_2004, lesgourges_2012, crotty_2004}. These approximations are widely used in cosmological analyses.

In this work, we obtain constraints  on the sum of neutrino masses for different mass hierarchies (NH, IH, DH and 1M), considering certain DE parameterizations - $\Lambda$CDM, Chevallier-Polarski-Linder (CPL) and exponential (EXP). Our aim is to see how neutrino mass upper bounds react to different parameterizations under different neutrino mass hierarchies. We also consider the effect of including different SNe Ia datasets (PantheonPlus and DESY5) on the total neutrino mass. We consider the EXP parameterization, which extends the CPL parameterization by introducing second-order corrections in redshift. This improves the flexibility of the CPL form while keeping the dimensionality fixed. Aside from using Bayesian inference (based on MCMC posteriors), we also use frequentist analysis to determine neutrino mass bounds since Bayesian parameter estimates could have  a strong dependence on the prior used~\cite{Patel, marc_2025, 2022_valent,Herold22}. This work was motivated by~\cite{marc_2025}. We have extended the analysis done in~\cite{marc_2025} to CPL and EXP parameterizations and by including SNe dataset and the recent DESI DR2 results. In this work, we do not limit ourselves to the impact of hierarchies but also the effect of DE parameterizations and the datasets used.

The paper is organized as follows. In Section~\ref{sec:2}, we describe how neutrinos affect cosmological observables and the DE parameterizations we use in this work. Section~\ref{sec:3} contains a description of the datasets and codes used. In Section~\ref{sec:4}, we state our results which has been further subdivided based on the focus of our comparison. Finally, Section~\ref{sec:5} provides our conclusions.

\section{Cosmological role of neutrinos and dark energy parameterizations}
\label{sec:2}

\textbf{Neutrinos}: Neutrinos leave distinct signatures on cosmological observables such as the Baryon Acoustic Oscillation (BAO) scale~\cite{seo_2003, blake_2003} and the growth of large-scale structure. In the early universe, weak interactions produced large amount of neutrinos, which decoupled from the primordial plasma at a temperature of about 1–2 MeV, while still being relativistic. After decoupling, neutrinos continued to freely stream through the universe, contributing to the total radiation energy density~\cite{komatsu_2011}. Their presence affects key cosmological epochs: during Big Bang Nucleosynthesis (BBN), neutrinos influence the expansion rate and thereby impact the abundance of light elements~\cite{steigman_2006, barger_2003,cyburt_2005}; during recombination, they alter the expansion history and the phase and amplitude of acoustic peaks in the Cosmic Microwave Background (CMB) power spectrum~\cite{bashinsky_2004, baumann_2016}. As the universe expands and cools, neutrinos lose energy due to cosmological redshift. Massive neutrinos become non-relativistic after the epoch of recombination, so that the onset of matter domination is not hastened and the early integrated Sachs-Wolfe effect~\cite{hu_1997, wayne_1996} is not  significantly altered~\cite{hou_2014, planck2_2020}. Hence, before recombination, massive neutrinos affect the radiation density while after recombination they contribute to the total matter density ($\Omega_m$). 

In the regime where the ISW effect remains unaffected by neutrinos, they affect cosmological observables by altering distances to the surface of last-scattering and suppress the growth of structure (late-time)~\cite{loverde_2024}. Massive neutrinos (due to their contribution towards $\Omega_m$) can decrease the angular diameter distance to last scattering $D_A=\int_0^{z_\text{rec}}\frac{dz}{H(z)}$. Since CMB constrains $\theta_s = \frac{r_s}{D_A}$ very precisely and $r_s$ (the sound horizon) is mainly dependent on pre-recombination physics, there needs to be a corresponding decrease in $H_0$. Neutrinos, due to their large velocity dispersions, reduce the clustering at scales smaller than their free-streaming scale ($k_{fs}$)~\cite{hu_1998, bond_1980}. Hence, we observe a suppression in the matter power spectrum at $k>k_{fs}$. This leads to a suppression in the CMB lensing signal~\cite{valentino_2024}. 

BAO measurements are made when the neutrinos are non-relativistic ($z<z_d$, where $z_d$ is the redshift of the baryon drag epoch). They constrain $r_dh$ and $\Omega_m$~\cite{ruchika_2024}. However, BAO is not \rthis{sensitive} enough to determine how the matter abundance is partitioned among baryons, cold dark matter and neutrinos. They also require an independent calibration of $r_d$ to infer the physical matter densities. This is where CMB comes in. CMB data helps calibrate the physical baryon and cold dark matter densities and the excess density is attributed to massive neutrinos (because before recombination, neutrinos contributed to the radiation density and so $\Omega_m$ only had baryonic and cold dark matter density contributions). These constraints are strengthened by CMB lensing, which is sensitive to the suppression in small-scale power caused by massive neutrinos \cite{giusarma_2018, tanseri_2022}.

There is a well-known degeneracy between $M_\text{tot}$ and the DE EOS~\cite{hannestad_2005, zhang_2017, shouvik_2018, yang_2017}. Since SNe data  help constrain the background expansion history — in particular $\Omega_m$ — it provides additional leverage to constrain $M_\text{tot}$.

\textbf{Dark Energy Parameterization}: In this work, we use three DE parameterizations defined by the EOS $w(z) \equiv  \frac{P(z)}{\rho(z)}$. The expansion history of the universe can be written as \cite{jheng_2024}:
\begin{equation}
    H^2(z) = H^2(0)\left[\Omega_r(1+z)^4 + \Omega_m(1+z)^3 + f_{DE}(z) + \Omega_k(1+z)^2\right],
\end{equation}
where $\Omega_r$, $\Omega_m$, and $\Omega_k$ represent the present dimensionless radiation, matter and curvature densities of the universe., respectively;  $f_{DE}$ is the term representing the DE dynamics given by:
\begin{equation}
    f_{DE}(z) = \Omega_{DE}\text{exp}{\left(3 \int_0^z \frac{1+w(z')}{1+z'}dz'\right)},
\end{equation}
where $\Omega_{DE} = 1 - (\Omega_r+\Omega_m+\Omega_k)$. In this work, we consider a spatially-flat universe, so $\Omega_k = 0$ henceforth.

\begin{itemize}
    \item \textbf{Standard $\Lambda$CDM}: Here, the DE EOS is constant and given by $w(z) = -1$.
    \item \textbf{CPL}: This is one of the most widely used DE parameterizations~\cite{chevallier_2001, linder_2003}. The EOS is given by $w(z) = w_0+\frac{w_az}{1+z}$. Advantages of the CPL parameterization is that it has a 2-dimensional phase space, exhibits a linear behaviour in redshift and has a straightforward physical interpretation, viz. $w_0$ corresponds to the dark energy EOS at the present time ($w_0 = w(z=0)$) while $w_a$ quantifies its redshift evolution.
    \item \textbf{EXP}: The exponential parameterization~\cite{pan_2020, najafi_2024, dimakis_2016, wolf_2025} has the following form for the EOS - $w(z) = (w_0-w_a)+w_a\text{exp}{\left(\frac{z}{1+z}\right)}$. Following~\cite{najafi_2024, giare_2024}, we consider terms upto second order in redshift such that the $w(z) = w_0 + w_a\left[\frac{z}{1+z}+\frac{1}{2!}\left(\frac{z}{1+z}\right)^2\right]$. Features of this parameterization are that it reduces to the CPL form when considering terms up to first order and it also considers corrections to the CPL parameterization without increasing the dimensionality of the parameter space. One of the objectives of this work is to see how the neutrinos mass constraints are affected by considering these corrections. A detailed analysis using the full exponential form is left for future investigation. 
\end{itemize}

\section{Data Sets and Methodology}
\label{sec:3}

\begin{itemize}
    \item \textbf{BAO}: We consider the BAO data from DESI DR2~\cite{desi_2025} from different tracers in seven redshift bins. The measurements include $D_V/r_d$ (for BGS), $D_M/r_d$ and $D_H/r_d$ and the correlations between $D_M/r_d$ and $D_H/r_d$.  
    \item \textbf{CMB}: We utilize CMB data from Planck PR3~\cite{planck1_2020, planck2_2020} which includes temperature and polarization (TT, TE, EE) and lensing power spectra. The low-$\ell$ likelihoods ($2 \leq \ell < 30$) are based on the \texttt{Commander} (for TT) and \texttt{SimAll} (for EE) likelihoods. The high-$\ell$ likelihoods ($30 \leq \ell \leq 2508$ for TT and $30 \leq \ell \leq 1996$ for TE and EE) use the \texttt{Plik} likelihood. The CMB lensing likelihood is based on the lensing reconstruction derived from the \texttt{SMICA} component-separated map. We do not use CMB lensing measurements from the more recent Atacama Cosmology Telescope (ACT) DR6~\cite{act1_2024, act2_2024} and Planck PR4~\cite{planck_pr4_2022}, since it requires an increase in precision, which subsequently slows down computations and for the purposes of this work is not relevant~\cite{marc_2025}. Henceforth, we shall refer to this combination of CMB dataset as `Planck' or `CMB'. Note that different Planck pipelines~\cite{hl_2024, camspec_2022} can lead to different upper bounds for neutrino mass constraints~\cite{desi_2024, desi_2025, elbers_2025, addison_2024, allali_2024}.
    \item \textbf{Type Ia SNe}: We use uncalibrated PantheonPlus (PP)~\cite{scolnic_2022, brout_2022} and Dark Energy Survey (DES) Y5~\cite{des_2024} SNe samples. PantheonPlus is a compilation of 1701 light curves of 1550 Type Ia SNe. We consider $z\geq0.01$ to remove any strong peculiar velocity dependencies. DESY5 is a compilation of 1829 Type Ia SNe (1635 in the range $0.1 < z < 1.13$ and 194 external low-redshift sample spanning $0.025<z<0.1$).
\end{itemize}

The Boltzmann code \texttt{CLASS}~\cite{class_2011, class2_2011} is used for linear theory predictions, while \texttt{halofit}~\cite{halofit_2003, halofit_2012} is employed for non-linear corrections to the matter power spectrum. We modify \texttt{CLASS} to include the CPL and EXP DE parameterizations. To compute Bayesian posteriors, we use the MCMC sampler \texttt{MontePython}~\cite{montepython_2013, montepython_2019}, and the package \texttt{GetDist}~\cite{getdist_2019} for analysis and visualization. In order to make DE subdominant in the early universe, we apply the conditions $w_0+w_a<0$ (denoted by $w_0w_a$) for CPL and $w_0+\frac{3}{2}w_a<0$ (denoted by $w_0w_a2$) for EXP. The essential priors considered are listed in Table~\ref{table1}. The relativistic degrees of freedom has been set to $N_\mathrm{eff} = 3.046$. The chains are believed to have converged when the Gelman-Rubin criterion of $R-1 \lesssim 0.015$ is satisfied. The likelihoods for DESI DR2 and DESY5 (adapted to \texttt{MontePython}) can be found at \href{https://github.com/tkarwal/cosmo_likelihoods.git}{$\mathrm{https://github.com/tkarwal/cosmo\_likelihoods.git}$}~\cite{jhaveri_2025, laura_herold_2025}.

We compute the profile likelihoods (PL) using \texttt{pinc}~\cite{pinc_2025}, which employs simulated-annealing algorithm for minimization and is interfaced with \texttt{MontePython}. The profile likelihood ratio (LR) test statistic is given by 
\begin{equation}
\label{eqn4}
    t_{\mu}^{LR}(x) = -2\text{log}\left(\frac{\mathcal{L}_x(\mu, \hat{\hat{\nu}})}{\mathcal{L}_x(\hat{\mu}, \hat{\nu})}\right),
\end{equation}
where $\hat{\mu}$, $\hat{\nu}$ and $ \hat{\hat{\nu}}$ denote MLE of the parameter of interest, nuisance parameters and the ``conditional'' MLE for $\nu$ which is obtained by keeping $\mu$ fixed, respectively. 

From Wilk's theorem \cite{wilks_1938}, we know that in the limit of large datasets, the LR test statistic follows a $\chi^2$ distribution for 1 degree of freedom.  Hence, Eqn.~\ref{eqn4} can be interpreted as a change in $\chi^2$: 
\begin{equation}
    \Delta\chi^2(M_\text{tot}) = -2\text{log}\left(\frac{\mathcal{L}(M_\text{tot}, \hat{\hat{\nu}})}{\mathcal{L}(\hat{M}_\text{tot}, \hat{\nu})}\right),
\end{equation}
where in our case $M_\text{tot}$ is the parameter of interest. Since $M_\text{tot}$ lies close to its physical boundary ($M_\text{tot}$ = 0), Wilk's theorem can break down. 
To address this, the Feldman-Cousins construction~\cite{neyman_1937, wald_1943, feldman_1998, pinc_2025, Barua2025} is used, which is already incorporated in the \texttt{pinc} analysis script. 
We have also  used Feldman-Cousins prescription to deal with the physical lower limits of 0.059 eV and 0.11 eV, while implementing parameter estimation for  NH and IH, respectively, similar to ~\cite{pinc_2025}.

We acknowledge the fact that the frequentist coverage may be approximate in our case, since a full Neyman construction which is required to guarantee full coverage is computationally expensive given the complexity of the models and the large datasets used \cite{pinc_2025, laura_herold_2025}.

\begin{table}[htbp!]
\caption{Flat priors imposed on the parameters for Bayesian analysis.}
\label{table1}
\centering
    \begin{tabular}{c@{\hspace{1cm}}c}
    \hline
    \thead{Parameter} & \thead{Priors} \\
    \hline
    \hline
        $\omega_\text{cdm}$        & [$-\infty, \infty$] \\
        $100\,\omega_b$            & [$-\infty, \infty$] \\
        $h$                        & [$-\infty, \infty$] \\
        $\ln(10^{10}A_s)$          & [$-\infty, \infty$] \\
        $n_s$                      & [$-\infty, \infty$] \\
        $\tau$                     & [0.004, $\infty$] \\
        $w_0$                      & [-5, 2] \\
        $w_0w_a$                   & [-10, -0.001] \\
        $w_0w_a2$                  & [-10, -0.001] \\
        $M_B$                      & [-30, -10]   \\
        $M_\text{tot}$ (1M, DH)    & [0, 1]       \\
        $M_\text{tot}$ (IH) \textsuperscript{$\dagger$}        & [0.11, 1]    \\
        $M_\text{tot}$ (NH) \textsuperscript{$\dagger$}        & [0.059, 1]   \\
    \hline
    \end{tabular}
    \parbox{0.9\linewidth}{\raggedright\small%
\textsuperscript{$\dagger$} The bounds for $M_\text{tot}$ in the IH and NH cases are computed within \texttt{MontePython} using neutrino oscillation data from ~\cite{esteban_2017} and then passed to \texttt{CLASS}}.
\end{table}

\section{Results}
\label{sec:4}

In this section, we compare the neutrino mass upper bounds according to three criteria:
\begin{itemize}
    \item Section~\ref{sec:4a}: comparison of mass hierarchies for a given DE parameterization and dataset (CMB$+$BAO and/or SNe) combination.
    \item Section~\ref{sec:4b}: assessing the impact of datasets used for a given choice of the DE parameterization and hierarchy. 
    \item Section~\ref{sec:4c}: evaluation of the effect of varying the DE parameterization for a particular hierarchy and dataset combination.
\end{itemize}

Note that in the subsequent subsections, when comparing frequentist and Bayesian upper bounds on $M_\mathrm{tot}$, we impose the same physical lower limit in the frequentist approach as the priors used in the Bayesian approach for the hierarchy under consideration. Specifically, for 1M and DH hierarchies we take $M_\mathrm{tot} \geq 0$ eV while for NH and IH, we use $M_\mathrm{tot} \geq 0.059$ eV  and $0.11$ eV, respectively. In Tables~\ref{table2}, \ref{table3}, \ref{table4} and \ref{table5}, we could not obtain non-zero central intervals on $M_\mathrm{tot}$ at 95\% confidence, hence we report only the 95\% upper bounds. In Appendix~\ref{appendixA}, we note the constraints on parameters other than $M_\mathrm{tot}$ for all configurations considered.

\begin{table}[htbp!]
\caption{95\% neutrino mass upper bounds for 1M neutrino mass hierarchy for both frequentist and Bayesian approaches. For the frequentist approach, we report upper bounds while considering $M_\mathrm{tot}>0$ eV. In the parenthesis, we state the upper bounds when considering NH and IH physical lower bounds: 0.059 eV for NH and 0.11 eV for IH (similar to Ref.~\cite{marc_2025}). For Bayesian approach, we take $M_\mathrm{tot}>0$ eV as physical lower limit.}
\label{table2}
\centering
    \begin{tabular}{c@{\hspace{1cm}}c@{\hspace{1cm}}c@{\hspace{1cm}}c}
    \hline
    \thead{Model/Dataset} & \thead{$M_\text{tot}$[eV]\\ \small(Bayesian)} & \thead{$M_\text{tot}$ (NH/IH lower limit)[eV]\\ \small(Frequentist)} \\
    \hline
    \hline
    & & \\[0.5ex]  
    \textbf{CMB$+$BAO} & & \\[1.5ex]
    $\Lambda$CDM & $<0.070$ & $<0.042\ (0.098/0.142)$\\[1ex]
    CPL          & $<0.205$ & $<0.205\ (0.207/0.223)$\\[1ex]
    EXP          & $<0.205$ & $<0.211\ (0.212/0.228)$\\[1ex]
    
    \textbf{CMB$+$BAO$+$PantheonPlus} & & \\[1.5ex]
    $\Lambda$CDM & $<0.075$ & $<0.050\ (0.101/0.144)$\\[1ex]
    CPL          & $<0.133$ & $<0.107\ (0.142/0.177)$\\[1ex]
    EXP          & $<0.140$ & $<0.122\ (0.146/0.178)$\\[1ex]
    
    \textbf{CMB$+$BAO$+$DESY5} & & \\[1.5ex]
    $\Lambda$CDM & $<0.082$ & $<0.059\ (0.104/0.145)$ \\[1ex]
    CPL          & $<0.157$ & $<0.134\ (0.157/0.186)$\\[1ex]
    EXP          & $<0.155$ & $<0.145\ (0.160/0.187)$\\[1ex]
    & & \\[0.5ex]  
    \hline
    \end{tabular}
\end{table}


\begin{table}[htbp!]
\caption{Same as Table~\ref{table2} but for degenerate neutrino mass hierarchy. For Bayesian approach, we take $M_\mathrm{tot}>0$ eV as physical lower limit.}
\label{table3}
\centering
    \begin{tabular}{c@{\hspace{1cm}}c@{\hspace{1cm}}c@{\hspace{1cm}}c}
    \hline
    \thead{Model/Dataset} & \thead{$M_\text{tot}$[eV]\\ \small(Bayesian)} & \thead{$M_\text{tot}$ (NH/IH lower limit)[eV]\\ \small(Frequentist)} \\
    \hline
    \hline
    & & \\[0.5ex]  
    \textbf{CMB$+$BAO} & & \\[1.5ex]
    $\Lambda$CDM & $<0.073$ & $<0.054\ (0.100/0.141)$\\[1ex]
    CPL          & $<0.198$ & $<0.199\ (0.200/0.215)$\\[1ex]
    EXP          & $<0.203$ & $<0.202\ (0.203/0.219)$\\[1ex]
    
    \textbf{CMB$+$BAO$+$PantheonPlus} & & \\[1.5ex]
    $\Lambda$CDM & $<0.078$ & $<0.065\ (0.103/0.142)$\\[1ex]
    CPL          & $<0.129$ & $<0.117\ (0.141/0.172)$\\[1ex]
    EXP          & $<0.138$ & $<0.124\ (0.145/0.176)$\\[1ex]
    
    \textbf{CMB$+$BAO$+$DESY5} & & \\[1.5ex]
    $\Lambda$CDM & $<0.084$ & $<0.074\ (0.106/0.144)$\\[1ex]
    CPL          & $<0.143$ & $<0.138\ (0.153/0.180)$\\[1ex]
    EXP          & $<0.155$ & $<0.146\ (0.159/0.185)$\\[1ex]
    & & \\[0.5ex]  
    \hline
    \end{tabular}
\end{table}


\begin{table}[htbp!]
\caption{Same as Table~\ref{table2} but for normal neutrino mass hierarchy. For Bayesian approach, we take $M_\mathrm{tot}>0.059$ eV as physical lower limit.}
\label{table4}
\centering
    \begin{tabular}{c@{\hspace{1cm}}c@{\hspace{1cm}}c@{\hspace{1cm}}c}
    \hline
    \thead{Model/Dataset} & \thead{$M_\text{tot}$[eV]\\ \small(Bayesian)} & \thead{$M_\text{tot}$ (NH/IH lower limit)[eV]\\ \small(Frequentist)} \\
    \hline
    \hline
    & & \\[0.5ex]  
    \textbf{CMB$+$BAO} & & \\[1.5ex]
    $\Lambda$CDM & $<0.116$ & $<0.064\ (0.102/0.141)$\\[1ex]
    CPL          & $<0.209$ & $<0.204\ (0.201/0.215)$\\[1ex]
    EXP          & $<0.208$ & $<0.206\ (0.203/0.217)$\\[1ex]
    
    \textbf{CMB$+$BAO$+$PantheonPlus} & & \\[1.5ex]
    $\Lambda$CDM & $<0.115$ & $<0.081\ (0.106/0.142)$\\[1ex]
    CPL          & $<0.158$ & $<0.128\ (0.143/0.171)$\\[1ex]
    EXP          & $<0.165$ & $<0.138\ (0.147/0.173)$\\[1ex]
    
    \textbf{CMB$+$BAO$+$DESY5} & & \\[1.5ex]
    $\Lambda$CDM & $<0.118$ & $<0.090\ (0.109/0.144)$\\[1ex]
    CPL          & $<0.173$ & $<0.147\ (0.155/0.179)$\\[1ex]
    EXP          & $<0.174$ & $<0.153\ (0.160/0.183)$\\[1ex]
    & & \\[0.5ex]  
    \hline
    \end{tabular}
\end{table}


\begin{table}[htbp!]
\caption{Same as Table~\ref{table2} but for inverted neutrino mass hierarchy. For Bayesian approach, we take $M_\mathrm{tot}>0.11$ eV as physical lower limit.}
\label{table5}
\centering
    \begin{tabular}{c@{\hspace{1cm}}c@{\hspace{1cm}}c@{\hspace{1cm}}c}
    \hline
    \thead{Model/Dataset} & \thead{$M_\text{tot}$[eV]\\ \small(Bayesian)} & \thead{$M_\text{tot}$ (NH/IH lower limit)[eV]\\ \small(Frequentist)} \\
    \hline
    \hline
    & & \\[0.5ex]  
    \textbf{CMB$+$BAO} & & \\[1.5ex]
    $\Lambda$CDM & $<0.153$ & $<0.046\ (0.098/0.141)$\\[1ex]
    CPL          & $<0.227$ & $<0.210\ (0.202/0.213)$\\[1ex]
    EXP          & $<0.241$ & $<0.220\ (0.208/0.216)$\\[1ex]
    
    \textbf{CMB$+$BAO$+$PantheonPlus} & & \\[1.5ex]
    $\Lambda$CDM & $<0.154$ & $<0.058\ (0.102/0.143)$\\[1ex]
    CPL          & $<0.191$ & $<0.125\ (0.143/0.173)$\\[1ex]
    EXP          & $<0.194$ & $<0.129\ (0.147/0.176)$\\[1ex]
    
    \textbf{CMB$+$BAO$+$DESY5} & & \\[1.5ex]
    $\Lambda$CDM & $<0.157$ & $<0.068\ (0.106/0.145)$\\[1ex]
    CPL          & $<0.201$ & $<0.144\ (0.155/0.180)$\\[1ex]
    EXP          & $<0.197$ & $<0.153\ (0.160/0.183)$\\[1ex]
    & & \\[0.5ex]  
    \hline
    \end{tabular}
\end{table}


\subsection{Comparison of hierarchies}
\label{sec:4a}

\begin{figure}
    \centering
    \subfloat[Bayesian 1D posterior\label{1a}]{\includegraphics[width=0.25\textwidth,keepaspectratio]{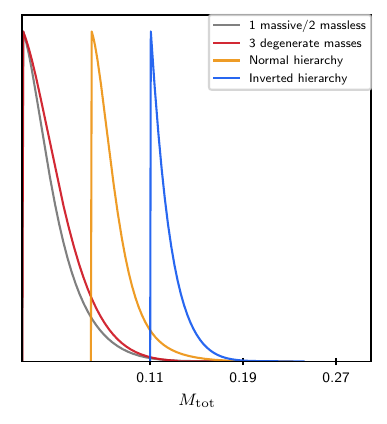}}
    \subfloat[Frequentist PL (Extrapolated to unphysical regime)\label{1b}]{\includegraphics[width=0.375\textwidth,keepaspectratio]{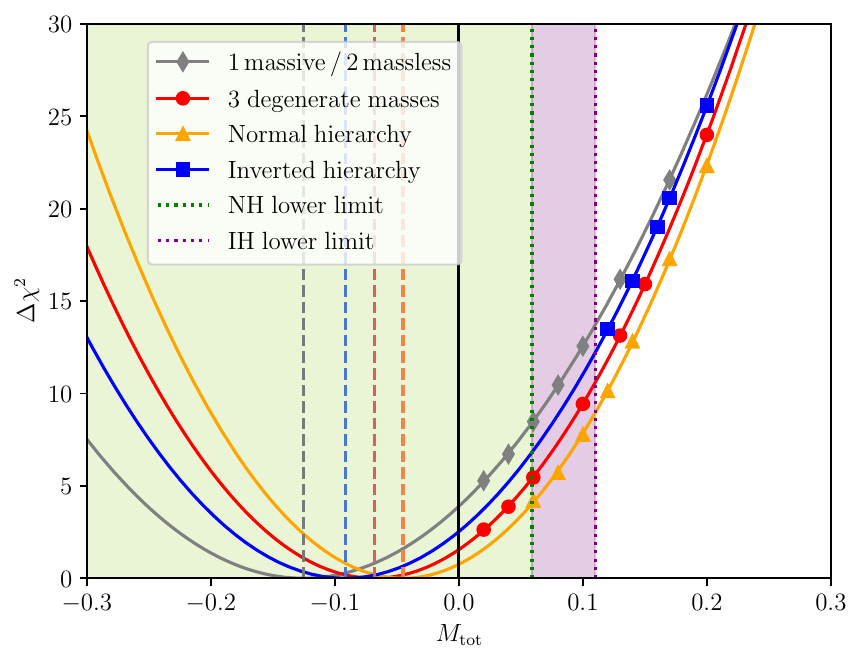}}
    \subfloat[Frequentist PL\label{1c}]{\includegraphics[width=0.375\textwidth,keepaspectratio]{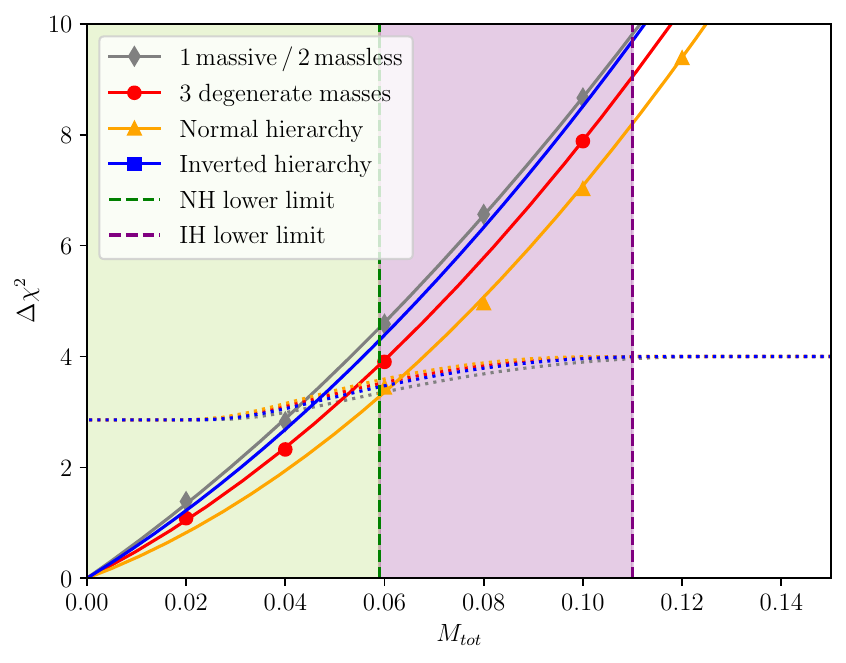}}
    \caption{Bayesian posteriors and frequentist PL for Planck$+$DESI $\Lambda$CDM comparing different hierarchies. Fig.~\ref{1a}: Bayesian 1D marginalized posterior for different mass hierarchies. Fig.~\ref{1b}: Frequentist PL for $M_\text{tot}$ extrapolated to unphysical region. The vertical dashed lines represent the minima of the parabolic curves. Fig.~\ref{1c}: Frequentist PL while considering $M_\mathrm{tot}>0$. Intersection of the curves with the dotted lines represent the 95\% upper bounds on $M_\mathrm{tot}$. $M_\mathrm{tot}$ has units of eV.}
    \label{fig1}
\end{figure}

\begin{figure}
    \centering
    \subfloat[Bayesian\label{2a}]{\includegraphics[width=0.25\textwidth,keepaspectratio]{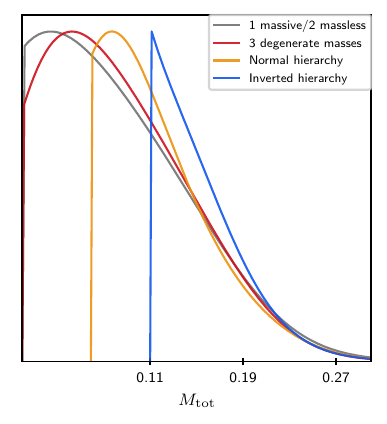}}
    \subfloat[Frequentist\label{2b}]{\includegraphics[width=0.375\textwidth,keepaspectratio]{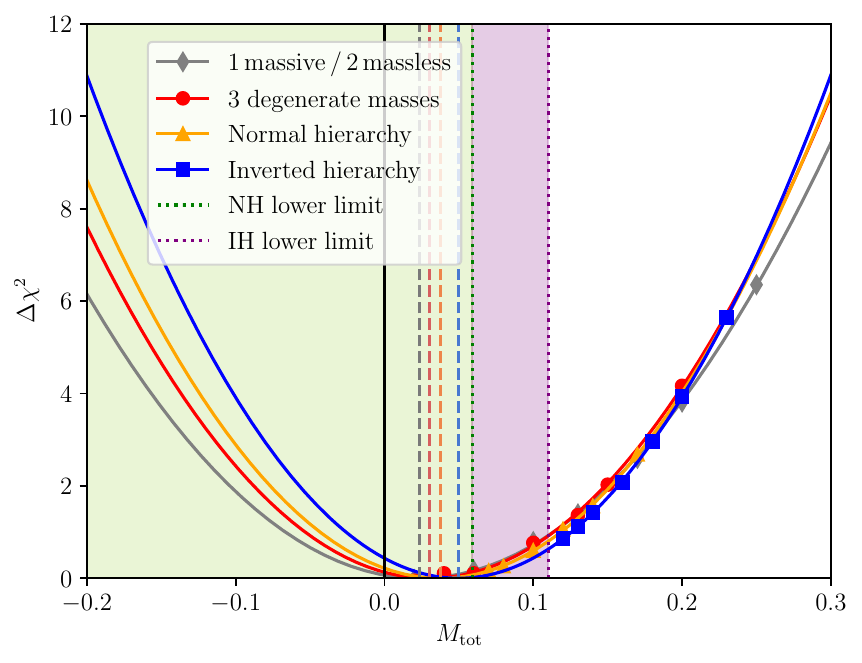}}
    \subfloat[Frequentist\label{2c}]{\includegraphics[width=0.375\textwidth,keepaspectratio]{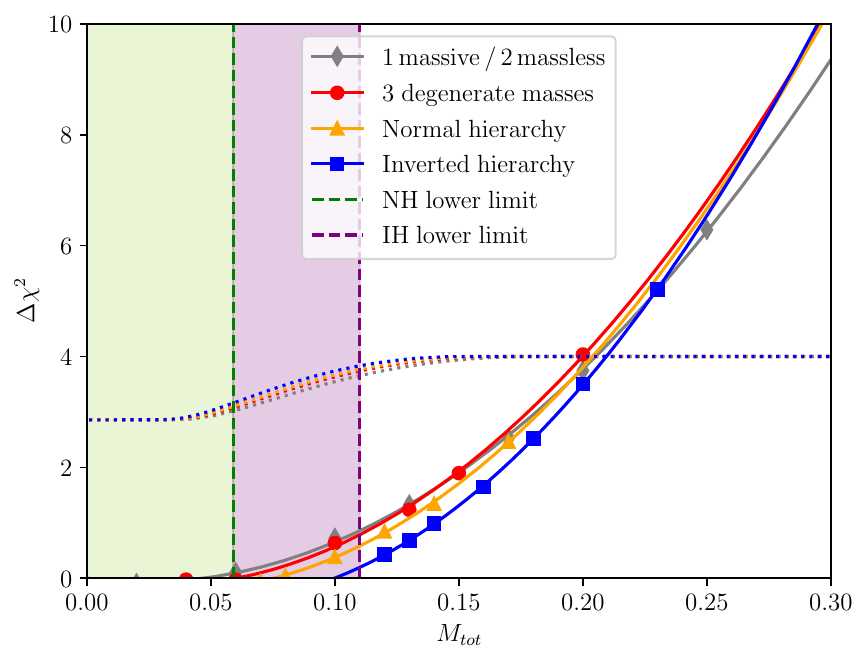}}
    \caption{Same as Fig.~\ref{fig1} but for Planck$+$DESI CPL. $M_\mathrm{tot}$ has units of eV.}
    \label{fig2}
\end{figure}

\begin{figure}
    \centering
    \subfloat[Bayesian\label{3a}]{\includegraphics[width=0.25\textwidth,keepaspectratio]{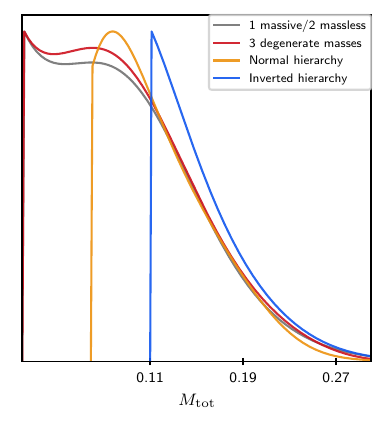}}
    \subfloat[Frequentist\label{3b}]{\includegraphics[width=0.375\textwidth,keepaspectratio]{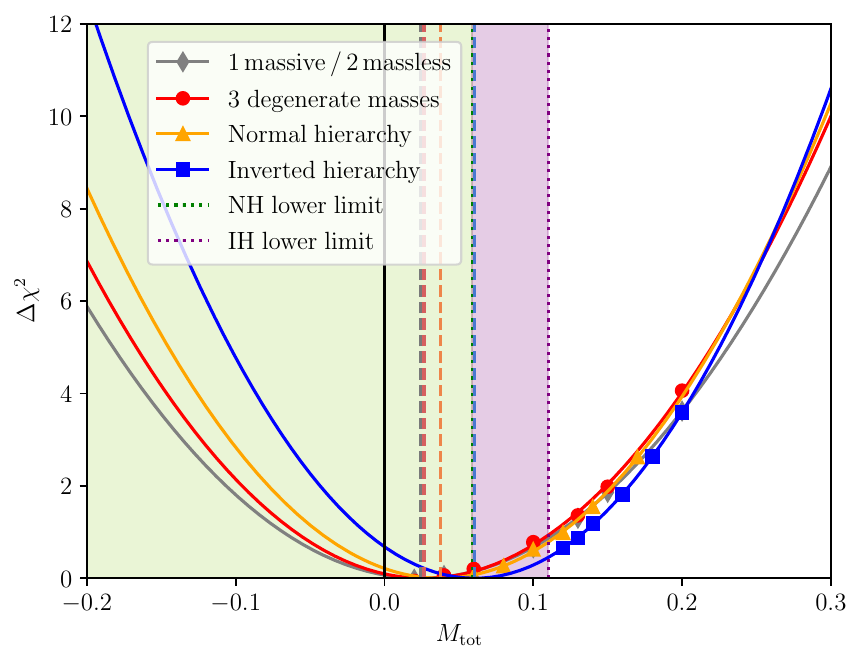}}
    \subfloat[Frequentist\label{3c}]{\includegraphics[width=0.375\textwidth,keepaspectratio]{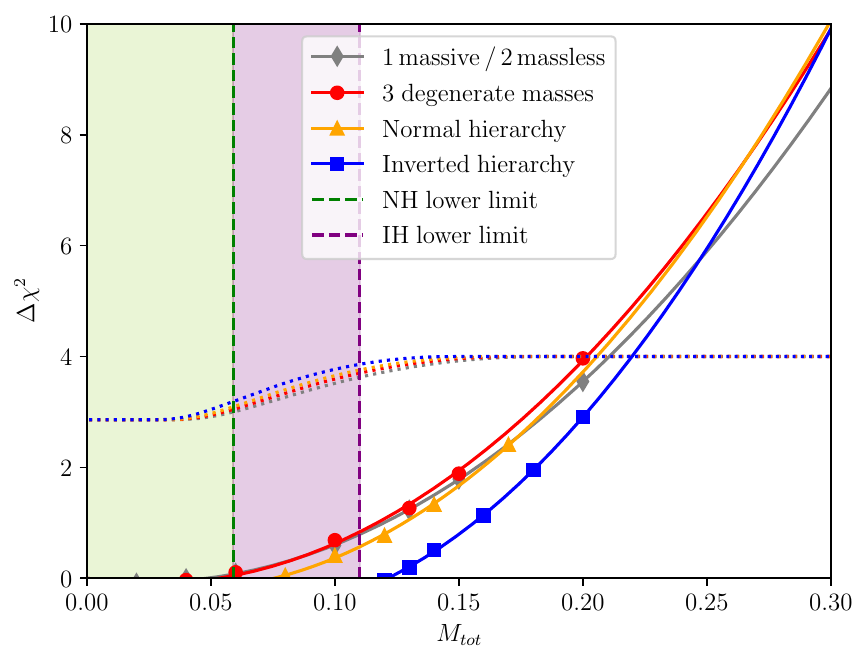}}
    \caption{Same as Fig.~\ref{fig1} but for Planck$+$DESI EXP. $M_\mathrm{tot}$ has units of eV.}
    \label{fig3}
\end{figure}

\begin{figure}
    \centering
    \subfloat[Bayesian\label{4a}]{\includegraphics[width=0.25\textwidth,keepaspectratio]{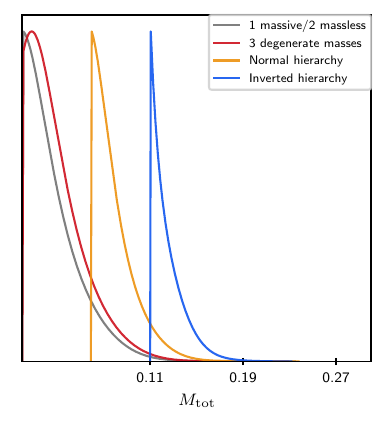}}
    \subfloat[Frequentist\label{4b}]{\includegraphics[width=0.375\textwidth,keepaspectratio]{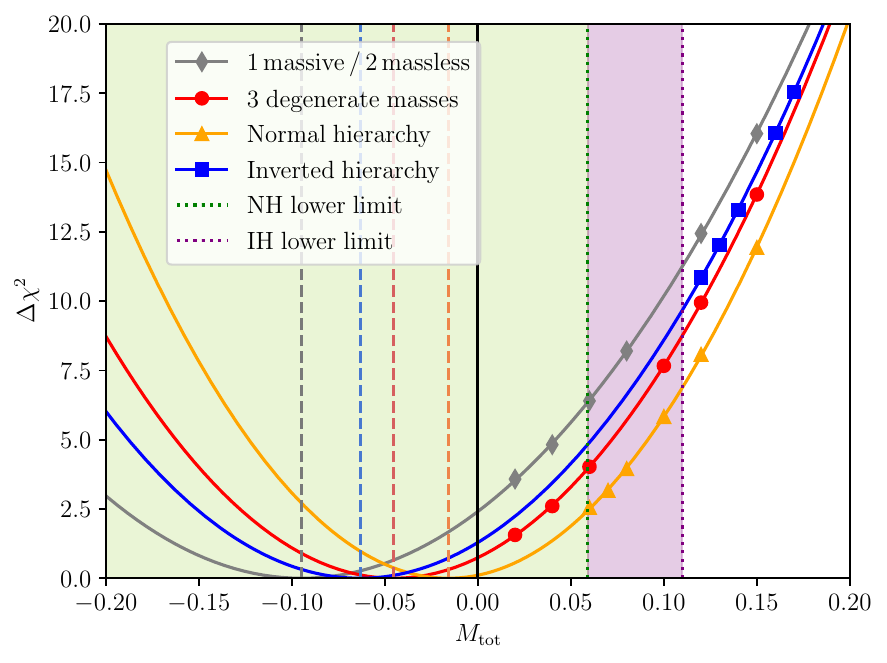}}
    \subfloat[Frequentist\label{4c}]{\includegraphics[width=0.375\textwidth,keepaspectratio]{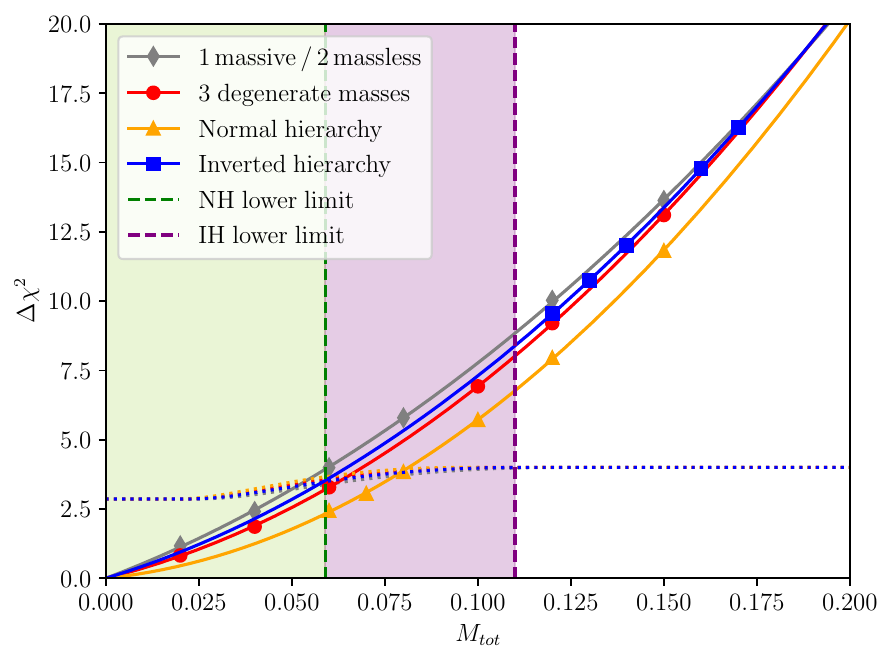}}
    \caption{Same as Fig.~\ref{fig1} but for Planck$+$DESI$+$PantheonPlus $\Lambda$CDM. $M_\mathrm{tot}$ has units of eV.}
    \label{fig4}
\end{figure}

\begin{figure}
    \centering
    \subfloat[Bayesian\label{5a}]{\includegraphics[width=0.25\textwidth,keepaspectratio]{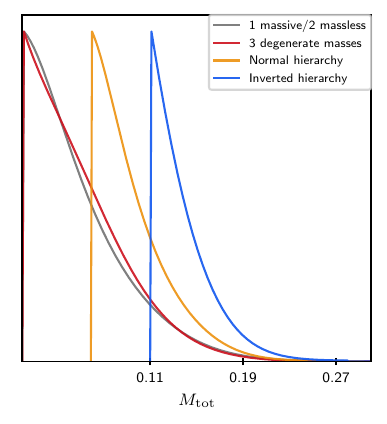}}
    \subfloat[Frequentist\label{5b}]{\includegraphics[width=0.375\textwidth,keepaspectratio]{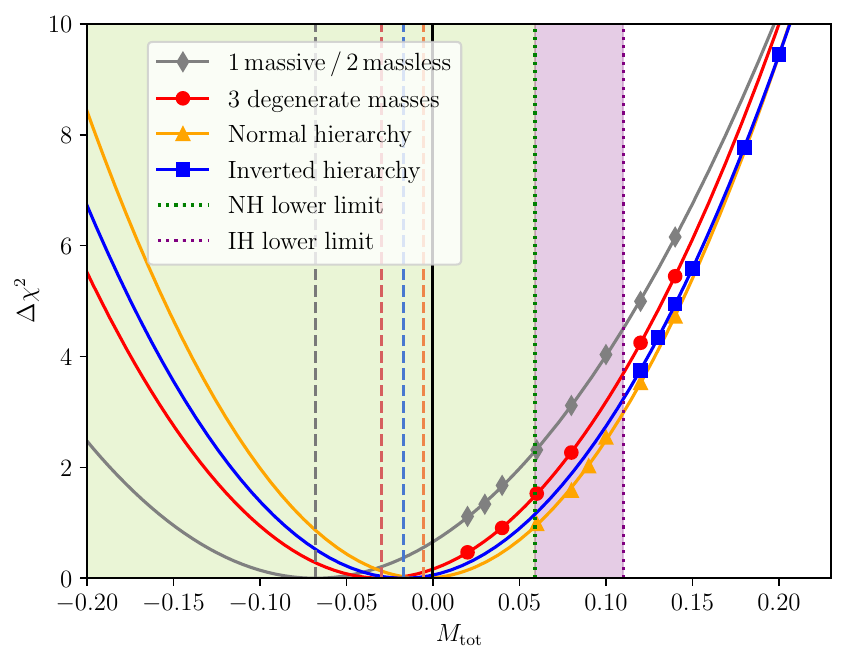}}
    \subfloat[Frequentist\label{5c}]{\includegraphics[width=0.375\textwidth,keepaspectratio]{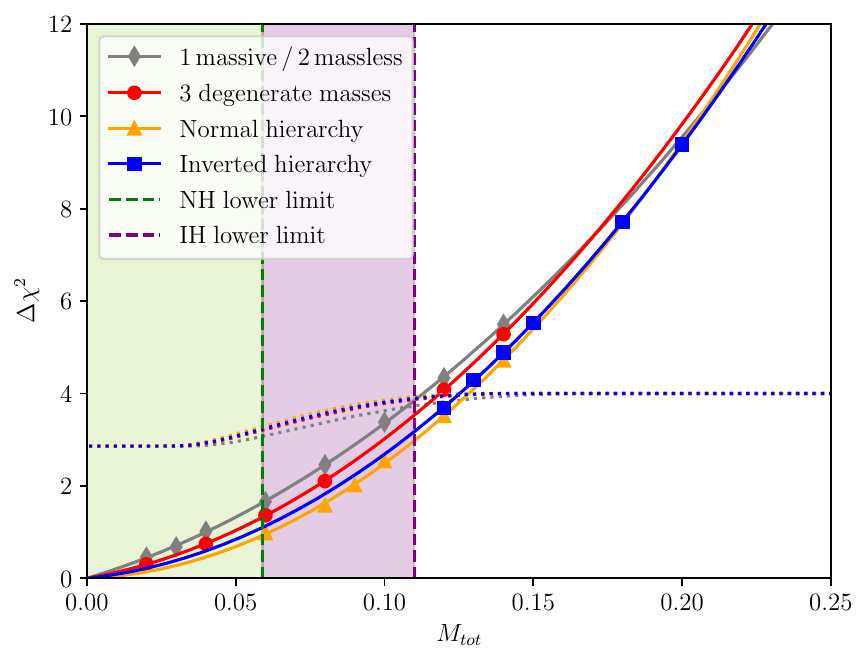}}
    \caption{Same as Fig.~\ref{fig1} but for Planck$+$DESI$+$PantheonPlus CPL. $M_\mathrm{tot}$ has units of eV.}
    \label{fig5}
\end{figure}

\begin{figure}
    \centering
    \subfloat[Bayesian\label{6a}]{\includegraphics[width=0.25\textwidth,keepaspectratio]{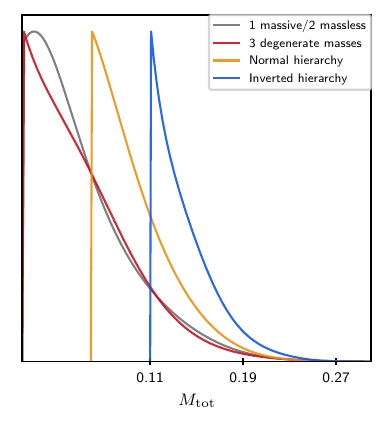}}
    \subfloat[Frequentist\label{6b}]{\includegraphics[width=0.375\textwidth,keepaspectratio]{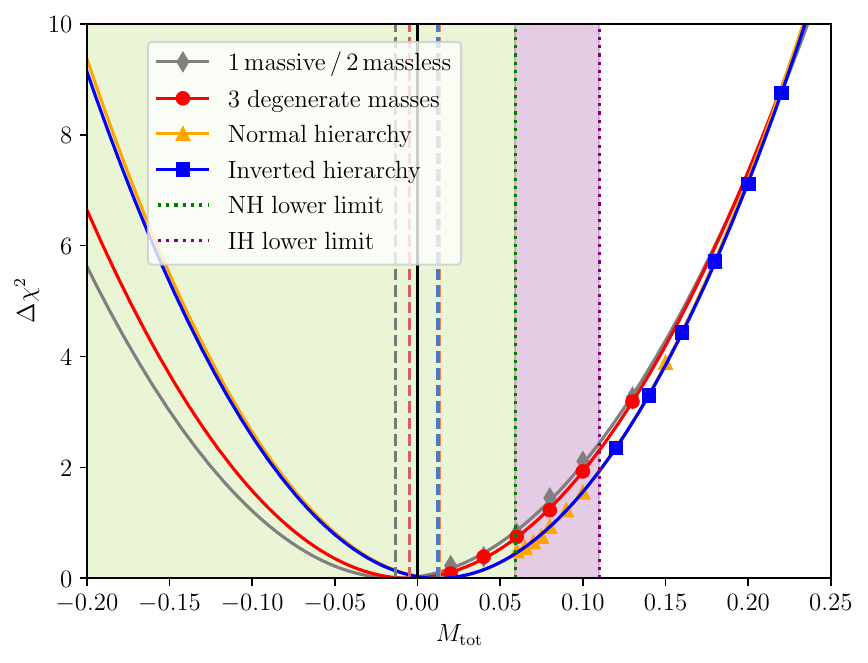}}
    \subfloat[Frequentist\label{6c}]{\includegraphics[width=0.375\textwidth,keepaspectratio]{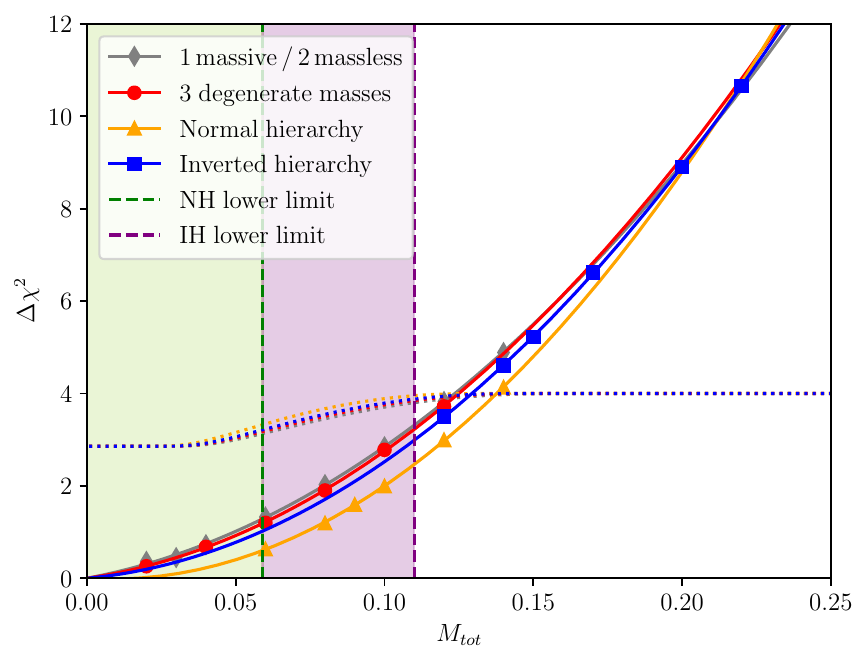}}
    \caption{Same as Fig.~\ref{fig1} but for Planck$+$DESI$+$PantheonPlus EXP. $M_\mathrm{tot}$ has units of eV.}
    \label{fig6}
\end{figure}

\begin{figure}
    \centering
    \subfloat[Bayesian\label{7a}]{\includegraphics[width=0.25\textwidth,keepaspectratio]{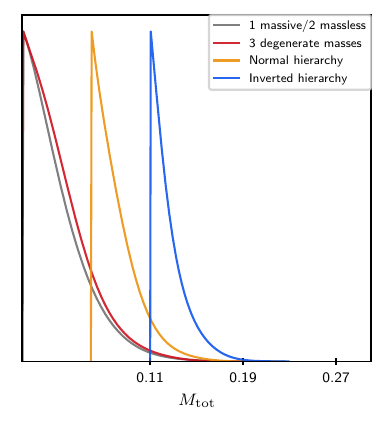}}
    \subfloat[Frequentist\label{7b}]{\includegraphics[width=0.375\textwidth,keepaspectratio]{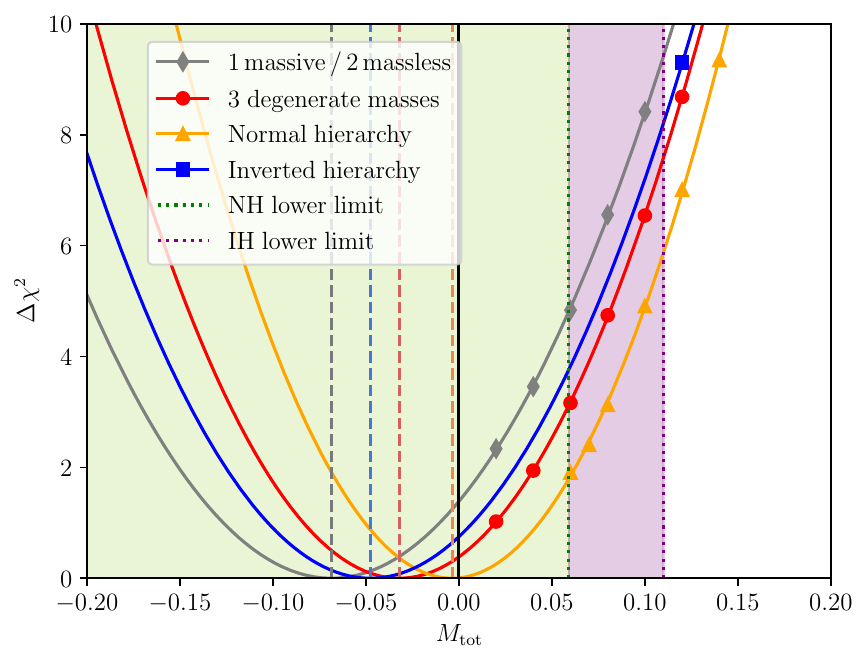}}
    \subfloat[Frequentist\label{7c}]{\includegraphics[width=0.375\textwidth,keepaspectratio]{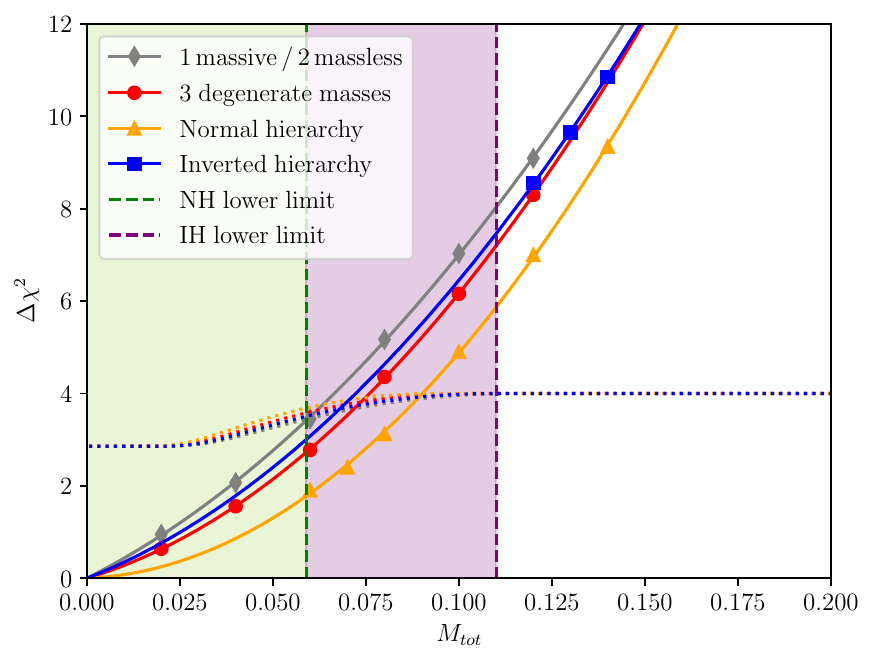}}
    \caption{Same as Fig.~\ref{fig1} but for Planck$+$DESI$+$DESY5 $\Lambda$CDM. $M_\mathrm{tot}$ has units of eV.}
    \label{fig7}
\end{figure}

\begin{figure}
    \centering
    \subfloat[Bayesian\label{8a}]{\includegraphics[width=0.25\textwidth,keepaspectratio]{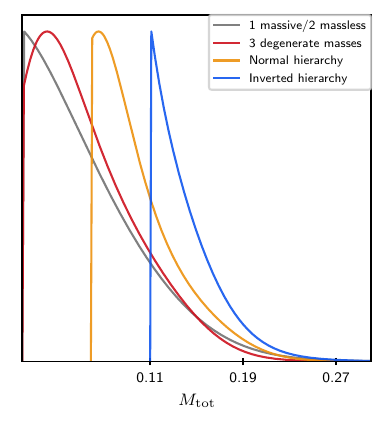}}
    \subfloat[Frequentist\label{8b}]{\includegraphics[width=0.375\textwidth,keepaspectratio]{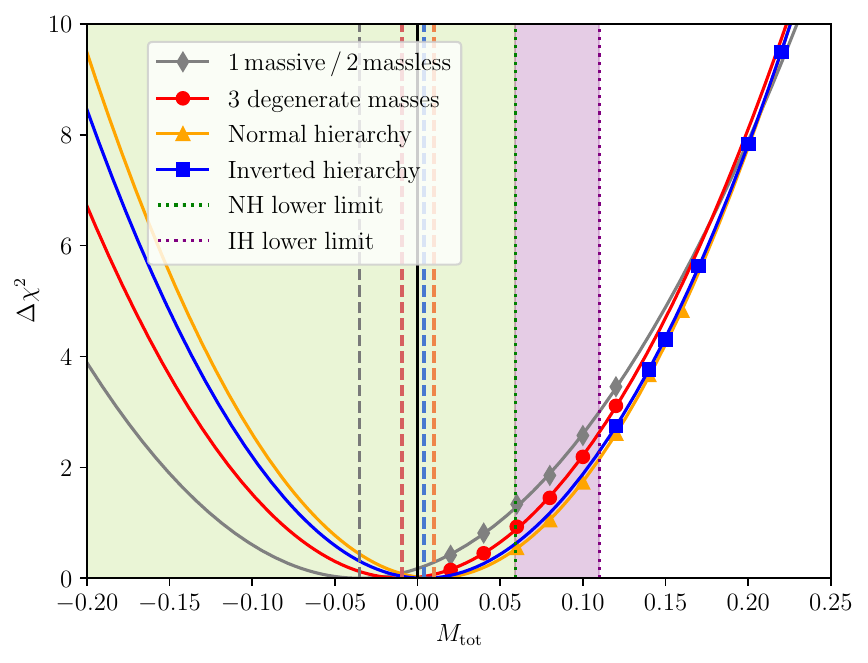}}
    \subfloat[Frequentist\label{8c}]{\includegraphics[width=0.375\textwidth,keepaspectratio]{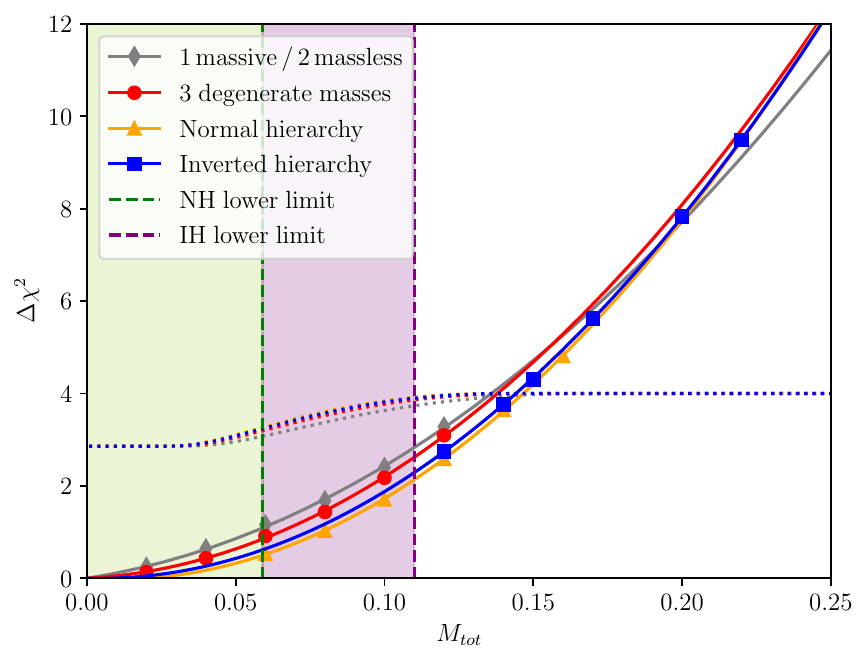}}
    \caption{Same as Fig.~\ref{fig1} but for Planck$+$DESI$+$DESY5 CPL. $M_\mathrm{tot}$ has units of eV.}
    \label{fig8}
\end{figure}

\begin{figure}
    \centering
    \subfloat[Bayesian\label{9a}]{\includegraphics[width=0.25\textwidth,keepaspectratio]{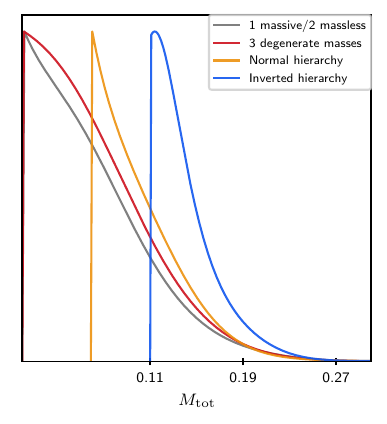}}
    \subfloat[Frequentist\label{9b}]{\includegraphics[width=0.375\textwidth,keepaspectratio]{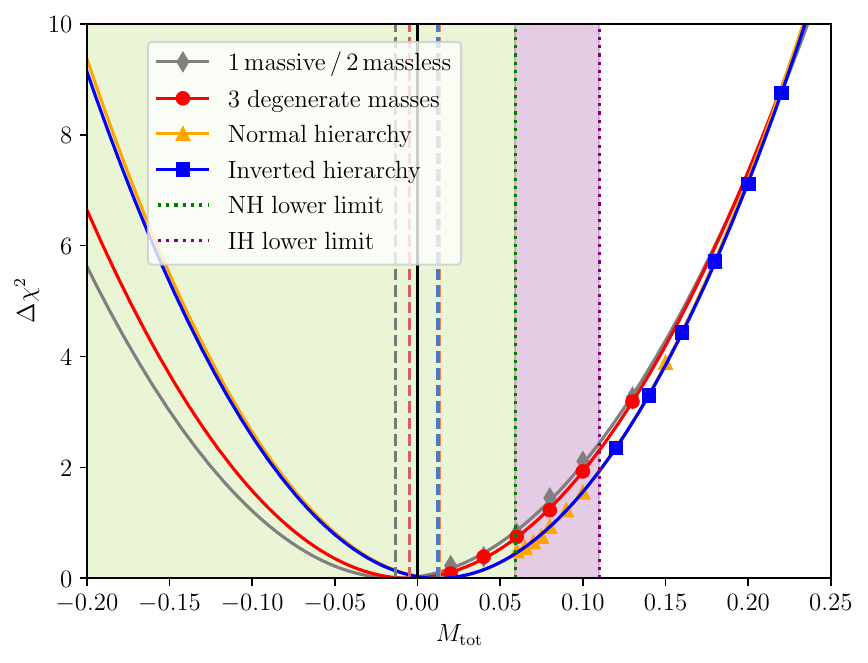}}
    \subfloat[Frequentist\label{9c}]{\includegraphics[width=0.375\textwidth,keepaspectratio]{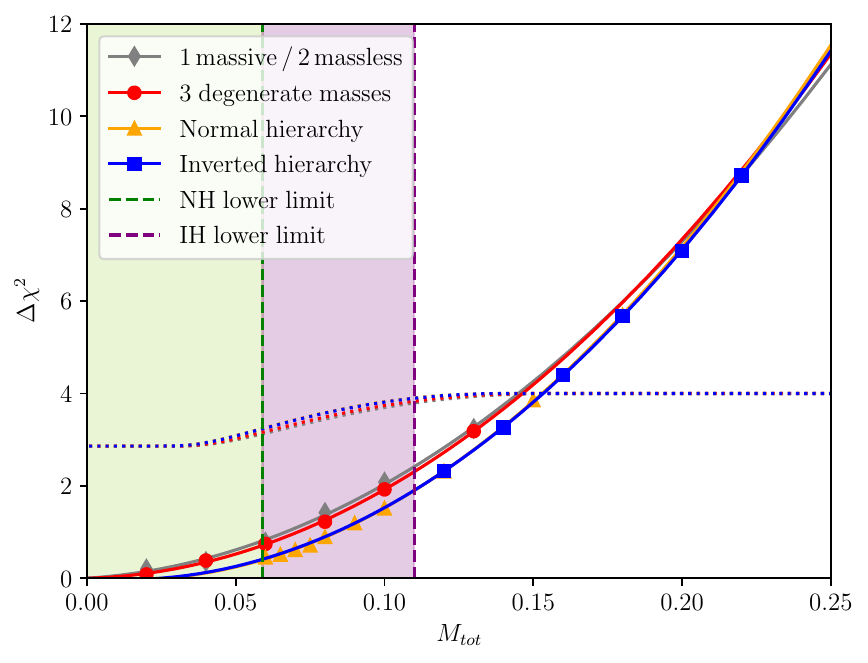}}
    \caption{Same as Fig.~\ref{fig1} but for Planck$+$DESI$+$DESY5 EXP. $M_\mathrm{tot}$ has units of eV.}
    \label{fig9}
\end{figure}

In this subsection, we consider the effect of hierarchies on a particular DE parameterization and dataset combination.
Refer to Figs.~\ref{fig1}-\ref{fig9} where we show the Bayesian posteriors and the frequentist PLs for a particular DE parameterization and dataset combination.
\begin{itemize}
    \item \textbf{CMB$+$BAO dataset} (Figs.~\ref{fig1}, \ref{fig2} and \ref{fig3}):
    \begin{itemize}
        \item For the $\Lambda$CDM model, the Bayesian 95\% upper bounds on the neutrino mass sum are relaxed by about 66\% (0.116 eV for NH) and 119\% (0.153 eV for IH) relative to the 1M (0.070 eV) case (with DH lying very close to 1M). In the CPL and EXP parameterizations, the Bayesian upper bounds are also relaxed but only by $\lesssim 15\%$. Considering the frequentist approach, the upper bounds from the $\Lambda$CDM model relax while going from 1M (0.042 eV) to DH (0.054 eV) to NH (0.064 eV) while the IH upper bound (0.046 eV) is similar to 1M. This trend is observed only when considering the lower limit of 0 eV for $M_\mathrm{tot}$. For CPL and EXP parameterizations, frequentist upper bounds do not show variations greater than 10\% for all choices of lower limits (0 eV, 0.059 eV,  and 0.11 eV). 
        \item Now, we compare Bayesian and frequentist upper bounds. For $\Lambda$CDM, the frequentist upper bounds are tighter than the Bayesian ones by $\sim(20-40)\%$ across all hierarchies. However, in the CPL and EXP parameterizations, the Bayesian upper bounds are relaxed by at most $10\%$ compared to the frequentist limits.
        \item For the EXP parameterization, the 1M and DH hierarchies seem to be less constrained by the data as can be seen from the flat posteriors in Fig.~\ref{fig3} (Bayesian case). 
        \item For $\Lambda$CDM the Bayesian posteriors do not show a peak in the positive regime of neutrino mass sum (Fig.~\ref{1a}). For both CPL and EXP parameterizations, the Bayesian posteriors (Fig.~\ref{2a}, \ref{3a}) for all hierarchies except IH show a peak which is cut-off at the boundary. 
        For 1M (0.070 eV) and DH (0.073 eV), considering $\Lambda$CDM, we find that the upper bounds on the neutrino mass sum are compatible with the physical limits 0 eV and 0.059 eV but not 0.11 eV. 
        The frequentist PLs for $\Lambda$CDM show a minima in the unphysical regime. For both CPL and EXP, we get minima in the region $M_\mathrm{tot}>0$. However, none of these minima are greater than the physical lower limits set by NH and IH oscillation experiments. The upper bounds inferred for $\Lambda$CDM (0.054 eV for 1M, 0.054 eV for DH, 0.064 eV for NH and 0.046 eV for IH) while considering $M_\mathbf{tot} > 0$ implies incompatibility with the physical lower limits of 0.059 eV and/or 0.11 eV. On the contrary, CPL and EXP show positive neutrino mass indications since they yield  larger upper bounds on $M_\mathrm{tot}$.
    \end{itemize}
    \item \textbf{CMB$+$BAO$+$PantheonPlus dataset} (Figs.~\ref{fig4}, \ref{fig5} and \ref{fig6}):
    \begin{itemize}
        \item In the Bayesian approach, the 1M and DH hierarchies yield identical bounds on $M_\mathrm{tot}$ within each parameterization---$\Lambda$CDM (0.075 eV, 0.078 eV), CPL (0.133 eV, 0.129 eV), and EXP (0.140 eV, 0.138 eV)---but differ across models. These bounds become more relaxed when moving to NH---53\% (0.115 eV) in $\Lambda$CDM, 18\% (0.158 eV) in CPL and 17\% (0.165 eV) in EXP---and to IH---105\% (0.154 eV) in $\Lambda$CDM, 43\% (0.191 eV) in CPL and 38\% (0.194 eV) in EXP. In the frequentist approach, with the lower boundary set to zero, all three DE parameterizations show an increase in the upper bound when going from 1M to DH to NH. For IH, however, the bound becomes more stringent---by less than $5\%$ in CPL/EXP and by about $40\%$ in $\Lambda$CDM---compared to NH. When we consider the experimentally inferred lower limits (0.059 eV and 0.11 eV), the variation in the bound is  very little of around  $(5-10)\%$.
        \item Comparing the Bayesian and frequentist upper bounds, we find that for $\Lambda$CDM, the Bayesian limits are relaxed by about $10$--$30\%$ relative to the frequentist ones. For CPL and EXP, a similar relaxation occurs but only by about $10$--$20\%$.
        \item The Bayesian posteriors are well constrained (no flat part) unlike the CMB$+$BAO EXP case (as mentioned above).
        \item The Bayesian posteriors do not show any peaks in the positive neutrino mass sum region for any hierarchy considering a specific parameterization. The upper bounds on the neutrino mass sum for $\Lambda$CDM---0.075 eV for 1M and 0.078 eV for DH---are compatible with the physical limits 0 eV and 0.059 eV but not 0.11 eV. Upper bounds for NH (0.115 eV) and IH (0.154 eV) are compatible with non-zero neutrino mass. From the frequentist PLs, we find that the minima for all hierarchy-parameterization combinations lie below $M_\mathrm{tot} = 0$ except in the EXP case for NH and IH (Fig.~\ref{fig6}). However, none of these minima are greater than the physical lower limits set by NH and IH oscillation experiments. The upper limits, inferred considering $M_\mathrm{tot} > 0$ using $\Lambda$CDM, show that none of the hierarchies---0.050 eV for 1M, 0.065 eV for DH, 0.081 eV for NH and 0.058 eV for IH---show upper bounds consistent with the IH physical lower limit of 0.11 eV. In both Bayesian and frequentist frameworks, the upper bounds for CPL and EXP remain compatible with non-zero neutrino masses due to the relaxed constraints.
    \end{itemize}
    \item \textbf{CMB$+$BAO$+$DESY5 dataset} (Figs.~\ref{fig7}, \ref{fig8} and \ref{fig9}):
    \begin{itemize}
        \item In the Bayesian approach, $\Lambda$CDM shows a relaxation of the upper bounds as we move from 1M (0.082 eV) and DH (for which the bounds are similar to 1M) to NH (0.118 eV) and IH (0.157 eV). The EXP parameterization exhibits a similar trend. For CPL, the behaviour is same, except for a slight tightening of about $9\%$ in the DH (0.143 eV) case relative to 1M (0.157 eV). In the frequentist approach, the relaxation of upper bounds is most pronounced in $\Lambda$CDM, where they increase from 1M (0.059 eV) to NH (0.090 eV) but become tighter for IH (0.068 eV) compared to DH (10\%) and NH (28\%). This trend is also present for CPL and EXP but is not as pronounced (2\% and 6\%, respectively when compared to NH). This occurs only for $M_\mathrm{tot} \geq 0$ eV. Frequentist upper bounds, considering 0.059 eV and 0.11 eV as the lower limits vary by less than 5\% across all hierarchies.
        \item Comparison of the frequentist and Bayesian upper bounds on $M_\mathrm{tot}$ show that the frequentist limits are consistently more stringent than the Bayesian ones.
        \item The Bayesian posteriors are well constrained unlike the CMB$+$BAO EXP case.
        \item The Bayesian posteriors for $\Lambda$CDM exhibit peaks placed in the unphysical region ($M_\mathrm{tot}<0$). The DH case in the CPL parameterization, however, shows a peak in the positive regime. Any other hierarchy-model combination either does not show a peak or gets cut-off at the boundary. 1M (0.082 eV) and DH (0.084 eV) upper bounds for $\Lambda$CDM are incompatible with the IH physical lower limit. For the frequentist PL, the minima for $\Lambda$CDM shows preference for negative neutrino mass. However, in the CPL (Fig.~\ref{8b}) and EXP (Fig.~\ref{9b}) DE parameterizations, NH and IH show minima in the positive regime considering $M_\mathrm{tot} \geq 0$. We also see that these minima lie below the NH and IH physical lower limits from the oscillation experiments. Frequentist upper bounds for $\Lambda$CDM (0.059 eV for 1M, 0.074 eV for DH, 0.090 eV for NH and 0.068 for IH), considering $M_\mathrm{tot}>0$, are inconsistent with the IH physical lower limit. In both Bayesian and frequentist frameworks, the upper bounds for CPL and EXP remain compatible with non-zero neutrino masses due to the relaxed constraints.
    \end{itemize}
\end{itemize}

\subsection{Comparison of datasets}
\label{sec:4b}

In this subsection, we assess the impact of the dataset combinations (CMB$+$BAO, CMB$+$BAO$+$PantheonPlus and CMB$+$BAO$+$DESY5) for a particular DE parameterization and hierarchy.

\begin{itemize}
    \item \textbf{$\Lambda$CDM model}:
    \begin{itemize}
        \item In the Bayesian approach, the upper bounds relax by about (15-20)\% for CMB$+$BAO$+$DESY5 (0.082 eV for 1M and 0.084 eV for DH) compared to CMB$+$BAO (0.070 eV for 1M and 0.073 eV for DH) in 1M and DH cases, while for the other hierarchies the relaxation is within 10\%. In the frequentist approach, when we impose $M_\mathrm{tot} \geq 0$ eV, the upper limits are relaxed by (25-40)\% when including SNe. This relaxation is dependent on the choice of the SNe sample. The relaxation reduces to within 10\% when we consider either 0.059 eV or 0.11 eV as the physical lower limits.  
        \item The Bayesian upper bounds on $M_\mathrm{tot}$ are relaxed ($\sim (10-67)\%$) compared to the frequentist upper limits.
        \item All dataset combinations for a given hierarchy and model show peaks (Bayesian posteriors) and minima (frequentist PLs) in the negative region of $M_\mathrm{tot}$ (Figs.~\ref{fig1}, \ref{fig4} and \ref{fig7}).
    \end{itemize}
    \item \textbf{CPL model}:
    \begin{itemize}
        \item The inclusion of SNe dataset tightens the Bayesian upper bounds by about (10-40)\%, seen in all hierarchies. The choice of SNe sample also affects the bounds, with DESY5 giving relaxed upper bounds than PantheonPlus. Similar behaviour is also observed in the frequentist approach when imposing any of the three lower limits - 0 eV, 0.059 eV, or 0.11 eV. 
        \item When we compare Bayesian and frequentist upper bounds, we find that for CMB$+$BAO the 1M and DH limits are similar for both approaches. However, for NH and IH, frequentist approach provides stronger bounds. 
        \item In all cases, CMB$+$BAO show indications of positive neutrino mass as seen from the minima of the frequentist PLs but only for $M_\mathrm{tot} \geq 0$ eV (Fig.~\ref{fig2}). The CMB$+$BAO$+$DESY5 dataset also show positive minima for the sum of neutrino mass but only for NH and IH (Fig.~\ref{fig8}).  None, however, are consistent with the experimental oscillation lower limits.
    \end{itemize} 
    \item \textbf{EXP model}:
    \begin{itemize}
        \item In the Bayesian approach, the upper bounds show a trend similar to the CPL case - inclusion of SNe tightens the upper limits with the strength depending on the SNe sample. The frequentist approach exhibit similar behaviour.
        \item Comparison of Bayesian and frequentist upper bounds , we notice similar pattern as the CPL case.
        \item Similar to the CPL model, CMB$+$BAO show indications of positive neutrino mass for $M_\mathrm{tot} \geq 0$ eV (Fig.~\ref{fig3}). CMB$+$BAO$+$DESY5 also show positive neutrino mass minima for NH and IH (Fig.~\ref{fig9}). Unlike the CPL case (Fig.~\ref{fig5}), however, PantheonPlus with CMB$+$BAO also move towards a positive minima for neutrino mass in the NH case (Fig.~\ref{fig6}). None of the minima are consistent with the oscillation lower limits, although the CMB$+$BAO dataset comes very close to 0.059 eV (Fig.~\ref{fig3}).
    \end{itemize}
\end{itemize}


\subsection{Comparison of DE parameterization}
\label{sec:4c}

In this subsection, we look at the neutrino mass upper bounds for $\Lambda$CDM, CPL and EXP parameterizations for a particular dataset combination and hierarchy.

\begin{itemize}
    \item \textbf{CMB$+$BAO}:
    \begin{itemize}
        \item In the Bayesian approach, CPL and EXP parameterizations yield upper bounds which differ by less than 10\% for a given hierarchy. In contrast, the upper bounds for $\Lambda$CDM (0.070 eV for 1M, 0.073 eV for DH, 0.116 eV for NH and 0.153 eV for IH) are tighter compared to CPL (0.205 eV for 1M, 0.198 eV for DH, 0.209 eV for NH and 0.227 eV for IH) and EXP (0.205 eV for 1M, 0.203 eV for DH, 0.208 eV for NH and 0.241 eV for IH) parameterizations by about $\sim 65\%$ (1M and DH) to $\sim(35-45)\%$ in other hierarchies.
        A similar trend is observed in the frequentist approach as well as for all lower bound choices. However, the tightening of the upper bounds reduce when we consider physical lower limits corresponding to NH (0.059 eV) and IH (0.11 eV) (with IH being the lowest).
        \item For all DE parameterizations, frequentist upper bounds are tighter compared to the Bayesian ones. For CPL and EXP parameterization, this tightness is about 10\% while for $\Lambda$CDM in 1M and DH cases, frequentist limits are tighter by about (30-40)\%.
        \item CPL and EXP parameterization show positive neutrino mass for each hierarchy (as seen from the minima of the frequentist PLs). However, none of them are compatible with the physical lower limits set by oscillation experiments (Figs.~\ref{fig2} and \ref{fig3}).
    \end{itemize}
    \item \textbf{CMB$+$BAO$+$PantheonPlus}:
    \begin{itemize}
        \item The Bayesian upper bounds are tighter in $\Lambda$CDM ((20-80)\%) than in the CPL and EXP parameterizations. EXP upper limits are relaxed by $\sim10\%$ compared to CPL parameterization. Similar behavior is observed in the frequentist approach for all three choices of the lower limit.
        \item Now we compare Bayesian and frequentist upper bounds. The frequentist upper limits are tighter by at most 33\% (for $\Lambda$CDM 1M and DH) for all three DE parameterizations within each hierarchy.
        \item There is no sign of positive neutrino mass (minima from the frequentist PLs) in any of the cases (Figs.~\ref{fig4}, \ref{fig5} and \ref{fig6}) when we include PantheonPlus SNe except for the EXP parameterization for NH (Fig.~\ref{fig6}). 
    \end{itemize}
    \item \textbf{CMB$+$BAO$+$DESY5}:
    \begin{itemize}
        \item In the Bayesian approach, the trend remains the same - upper bounds from the CPL and EXP parameterizations are more relaxed compared to $\Lambda$CDM. Upper bounds from the EXP parameterization is 8\% more relaxed than in the CPL parameterization for DH. In all other hierarchies, there is no significant difference. A similar behavior is observed in the frequentist approach for all three choices of physical lower limits. EXP parameterization shows consistently relaxed upper bounds compared to CPL parameterization. 
        \item Comparing Bayesian and frequentist upper limits, we see $\Lambda$CDM, CPL and EXP provide tighter frequentist upper bounds compared to the Bayesian upper bounds.
        \item Compared to the PantheonPlus case above, we find that the frequentist PL minima lie in the positive neutrino mass region for both CPL (Fig.~\ref{fig8}) and EXP (Fig.~\ref{fig9}) parameterizations in NH and IH cases. However, none of them conform to the physical lower limits set by oscillation experiments.
    \end{itemize}
\end{itemize}


\section{Conclusions}
\label{sec:5}

\begin{figure}
    \centering
    \subfloat[1M\label{10a}]{\includegraphics[width=0.5\textwidth,keepaspectratio]{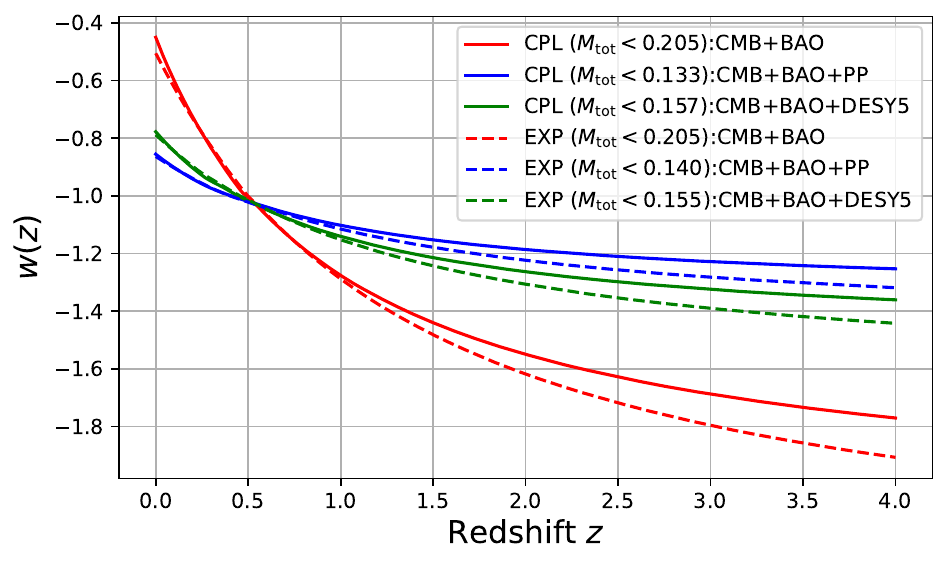}}
    \subfloat[DH\label{10b}]{\includegraphics[width=0.5\textwidth,keepaspectratio]{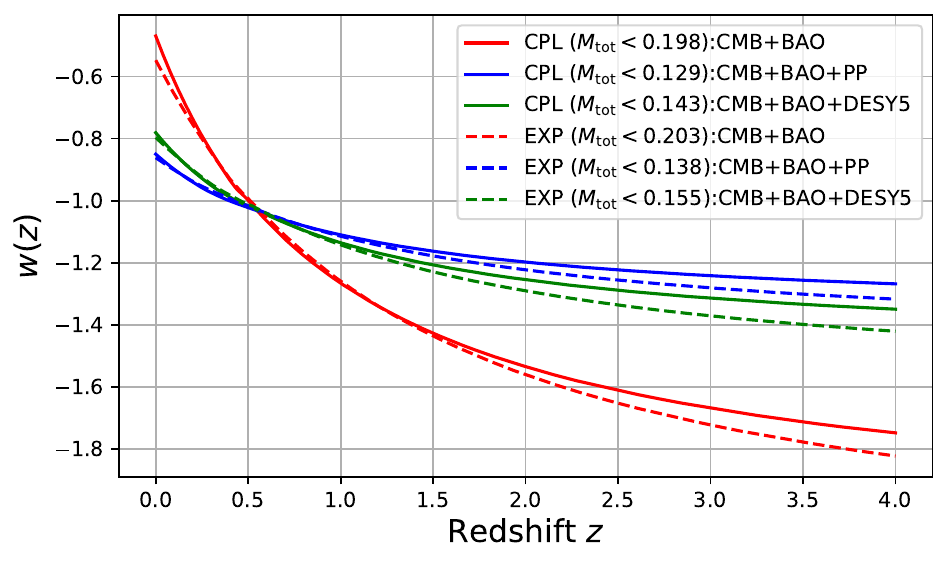}} \\
    \subfloat[NH\label{10c}]{\includegraphics[width=0.5\textwidth,keepaspectratio]{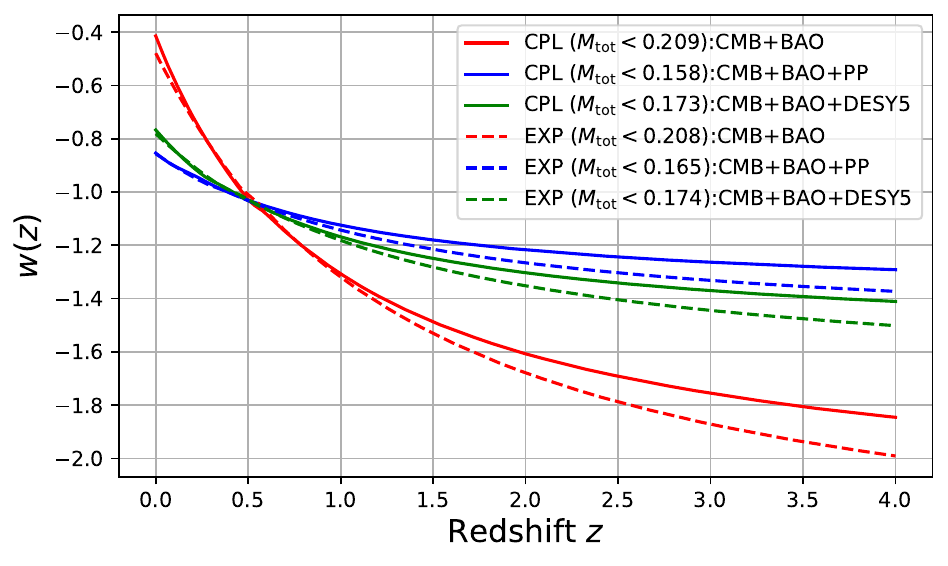}}
    \subfloat[IH\label{10d}]{\includegraphics[width=0.5\textwidth,keepaspectratio]{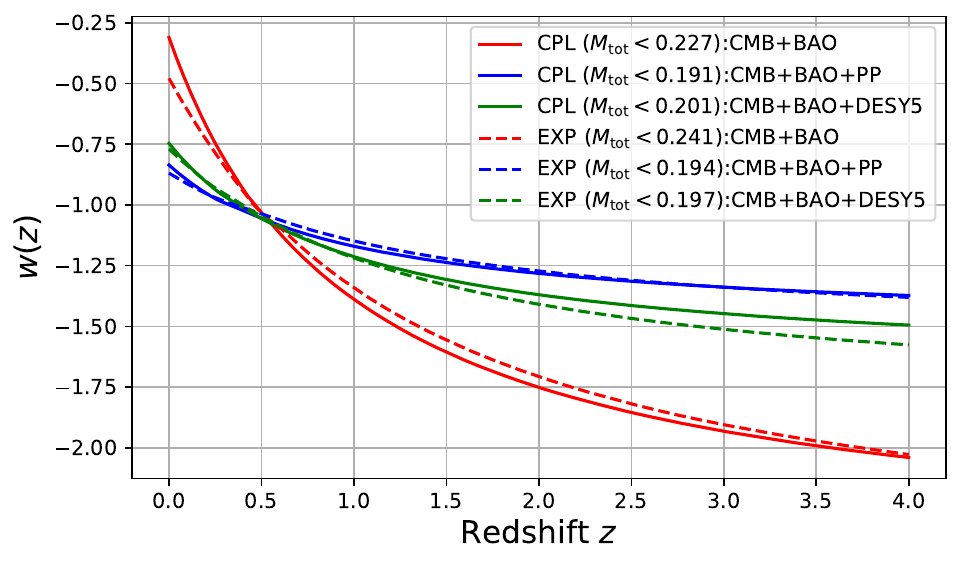}}
    \caption{The EOS $w(z)$ is shown as a function of redshift $z$ for both CPL and EXP parameterization . The EOS for CPL is shown in solid lines while for EXP is plotted using dashed lines. Comparison is made for a particular hierarchy and dataset combination and the Bayesian upper limits on $M_\mathrm{tot}$ (in eV) are indicated in the legend. For brevity, we omit the 68\% confidence band and show only the median EOS curve.}
    \label{fig10}
\end{figure}

\begin{figure}
    \centering
    \includegraphics[width=1.0\textwidth]{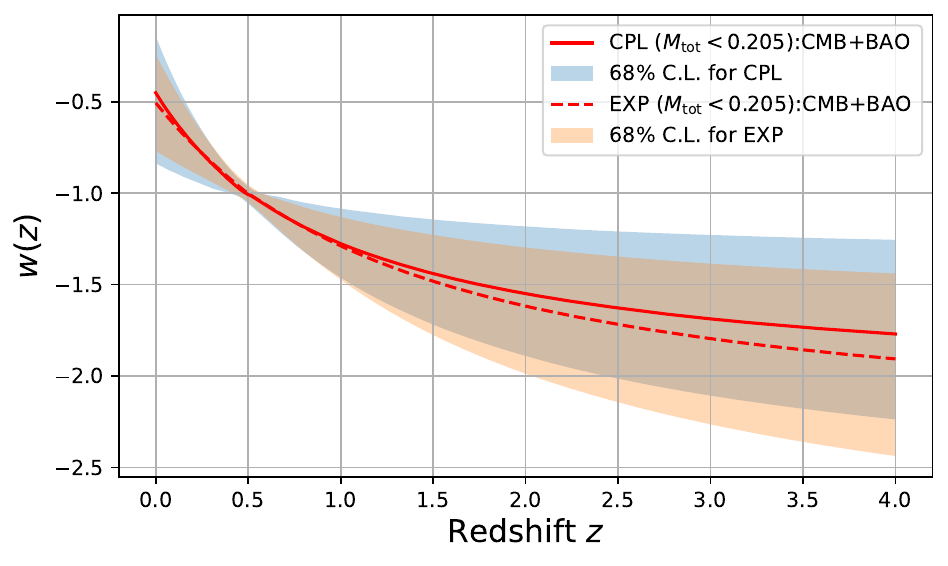}
    \caption{\rthis{The EOS is shown for both CPL and EXP parameterization for CMB$+$BAO dataset and 1M hierarchy. This figure is similar to Fig.~\ref{10a} but we have included 68\%  confidence intervals. This shows that the DE EOS for a specific hierarchy-dataset combination for CPL and EXP parameterizations are consistent within $1 \sigma$ (as can be seen from Tables~\ref{table6}-\ref{table9}). This is true for all other cases, and  hence we show only a particular case here.}}
    \label{fig11}
\end{figure}

In this work, we utilized Planck (PR3) CMB data, BAO data from DESI DR2, and SNe data from the PantheonPlus and DESY5 samples to constrain $\Lambda$CDM$+M_\mathrm{tot}$, CPL$+M_\mathrm{tot}$, and EXP$+M_\mathrm{tot}$, considering four neutrino mass hierarchies: one massive/two massless (1M), degenerate (DH), normal (NH), and inverted (IH). Our analyses were performed using both Bayesian and frequentist approaches. In the Bayesian case, we imposed lower limits on the prior for $M_\mathrm{tot}$: 0 eV for 1M and DH, and (motivated by neutrino oscillation experiments) 0.059 eV for NH and 0.11 eV for IH. In the frequentist analysis, we enforced these same lower limits but for each hierarchy. The objectives of our work were to determine the impact of adding SNe dataset on $M_\mathrm{tot}$ upper bounds, assess how different DE parameterizations affect the upper bounds on $M_\mathrm{tot}$ (specifically, what changes can be seen when extending CPL to include second order correction term as in EXP), and look at the effect of hierarchies  on the upper bounds for $M_\mathrm{tot}$. Additionally, we also look at the difference between the Bayesian and the frequentist upper bounds. 

The parameter constraints from different DE parameterizations, dataset and hierarchy combinations can be found in Tables~\ref{table6}, \ref{table7}, \ref{table8}, and \ref{table9}. The neutrino mass upper bounds from both frequentist and Bayesian frameworks can be found in Tables~\ref{table2}, \ref{table3}, \ref{table4}, and \ref{table5}. In Figs.~\ref{fig1}-\ref{fig9}, we show the Bayesian posteriors and frequentist PLs grouped by hierarchy for each dataset-model combination. 

When we add  SNe to the CMB$+$BAO dataset combination, we notice that both the Bayesian and frequentist upper bounds tighten for the CPL and EXP parameterization for each hierarchy (Tables~\ref{table2}-\ref{table5}). BAO constrains the late-time expansion history in $\Lambda$CDM quite well~\cite{desi_2024, desi_2025}. However, extensions of $\Lambda$CDM expand the late-time parameter space. So, the constraining power of BAO weakens. Including SNe improves the constraints and hence, we get  tighter bounds. DESY5 gives relaxed $M_\mathrm{tot}$ upper limits as compared to PantheonPlus. 
However, for $\Lambda$CDM, including SNe relaxes the upper bounds. Only for CPL and EXP in 1M and DH, the upper limits from Bayesian and frequentist techniques are similar for the CMB$+$BAO dataset combination. This is due to the role of SNe (as discussed previously) in constraining late-time expansion history. When not including SNe samples, the extended parameter space makes the upper bounds larger and hence, both approaches have less constraining power leading to an apparent agreement in the frequentist and Bayesian approaches.

When considering a particular dataset and hierarchy (Section~\ref{sec:4c}), we note that the  upper bounds from $\Lambda$CDM are tighter compared to both CPL and EXP parameterization. This is easily explained by the complexity of the DE EOS, which expands the parameter space~\cite{hannestad_2005, vagnozzi_2018, scr_2019, naredo_2024, elbers_willem_2025, desi_2024, desi_2025, elbers_2025}. We also observe that EXP gives slightly more relaxed upper limits ($\lesssim10\%$) than CPL (Tables~\ref{table2}-\ref{table5}). This is true for both frequentist and Bayesian approaches and is due to the additional second order term introduced compared to the popular CPL parameterization.

In Fig.~\ref{fig10} we show the EOS for CPL and EXP as a function of redshift. For clarity, we only plot the median EOS curve but note that $w(z)$ agrees within $1\sigma$ at all redshifts. This can also be seen from the $w_0$ and $w_a$ values in Tables~\ref{table6}-\ref{table9}. \rthis{We have also shown the 1M case for CMB$+$BAO while including the 68\% confidence band in Fig.~\ref{fig11}}. We note that at redshift of $z=0$, the parameterizations agree on $w (z=0)$ values for all dataset-hierarchy combinations (this is expected from the mathematical forms for both parameterizations). The very slight differences is due to the degeneracies present between $M_\mathrm{tot}$ and the other model parameters. We see that at redshifts $ z \gtrsim 0.5$, the EOS enters a phantom regime $(w(z) \lesssim -1)$. At higher redshifts, there is a deviation (within $1\sigma$) of the EOS due to the higher order terms in the EXP parameterization. This is the reason for the relaxed upper limits in the EXP as compared to the CPL parameterizations.

We notice that when we introduce SNe data, the EOS tends to become less negative compared to SNe-less data. This explains the tightening of the bounds in CMB$+$BAO$+$SNe data compared to SNe-less data. 
Fig.~\ref{fig10} also shows the dependence of the SNe compilation used on $M_\mathrm{tot}$ with PantheonPlus giving tighter constraints compared to DESY5 due to the EOS being slightly less negative for PantheonPlus compared to DESY5.

The above trends can be explained as follows: $\theta_s$ is constrained by CMB data quite accurately and $r_s$ is mainly dependent on pre-recombination physics. This leaves the angular diameter distance to last scattering ($D_A$) as a free parameter which needs to be fixed. We know that $D_A \propto \int_0^{z_\mathrm{rec}}\frac{dz}{H(z)}$. To keep it fixed, the parameters such as $\Omega_m$, $H_0$ and $M_\mathrm{tot}$ change their values according to the DE parameterization used~\cite{vagnozzi_2018, scr_2019, du_2025}.

For a particular choice of DE parameterization and dataset combination (Section~\ref{sec:4a}), Bayesian upper bounds are similar for 1M and DH but get relaxed for NH and IH. This can be attributed to the lower limits considered in the prior specification~\cite{marc_2025}. Frequentist upper limits (while considering $M_\mathrm{tot} \geq 0$) relax while going from 1M to DH to NH but become stringent for IH. This change is most prominent in $\Lambda$CDM as compared to CPL or EXP. Further, for the other lower limits (0.059 eV and 0.11 eV), the bounds are similar for all hierarchies ($(5-10)\%$ variation). Additionally, we also find that the bounds obtained from the frequentist approach depend on the lower limit imposed on $M_\mathrm{tot}$~\cite{simpson_2017, gariazzo_2018, gariazzo_2023, heavens_2018, vagnozzi_2017}. In all cases, we notice that the Bayesian upper bounds are more relaxed compared to the frequentist ones. This appears to be a product of the difference in the interpretations in the two statistical approaches. This point has also  been discussed in~\cite{marc_2025}.

As seen from the frequentist PL minima, for the $\Lambda$CDM model, none of the dataset–hierarchy combinations yield a positive neutrino mass (Figs.~\ref{fig1}, \ref{fig4}, and \ref{fig7}). In contrast, both CPL and EXP parameterizations show evidence of positive neutrino mass when using the CMB$+$BAO dataset (Figs.~\ref{fig2} and \ref{fig3}). Upon including SNe data, CPL no longer shows positive neutrino mass for PantheonPlus (Fig.~\ref{fig5}); however, the NH and IH hierarchies still indicate \rthis{evidence for non-zero} positive \rthis{neutrino} mass with the DESY5 sample (Figs.~\ref{fig8}). For EXP, positive neutrino mass is obtained for both SNe datasets (PantheonPlus and DESY5) when considering NH and IH (Figs.~\ref{fig6} and \ref{fig9}). However, none of these apparent detections remain once the physical lower limits from neutrino oscillation experiments ($M_\mathrm{tot} \geq 0.059$ eV for NH and $M_\mathrm{tot} \geq 0.11$ eV for IH) are imposed. Furthermore, (as discussed in Section~\ref{sec:4a}) the Bayesian upper bounds for the neutrino mass sum for $\Lambda$CDM do not show agreement with the IH-based physical lower limits \cite{jiang_2025}. The bounds do exceed 0.11 eV for NH but marginally. This indicates a tension with IH. On the other hand, CPL and EXP parameterizations, due to their relaxed bounds, show consistency with all three lower limits (0 eV, 0.059 eV and 0.11 eV). Even considering the frequentist upper limits (for $M_\mathrm{tot}>0$) lead to a tension between the upper bounds on the neutrino mass sum determined for $\Lambda$CDM with the physical lower limits set by the oscillation experiments. Our results show that whether or not we see signs of a positive neutrino mass depends on the choice of priors, the mass hierarchy (mildly) and the DE parameterization considered rather than only on the data itself. This is evident in both the Bayesian posteriors and the frequentist PL (extended to unphysical regimes).

We note that detections of positive neutrino mass at the $\gtrsim 2\sigma$ level have been found in literature, such as, $M_\mathrm{tot} = 0.19^{+0.15}_{-0.18}$ eV~\cite{shouvik_2025}. 
This, unlike our findings, are consistent with the lower limits obtained from the oscillation experiments. With this, we can confirm that the weak lensing data and and extended parameter space (including the $A_\mathrm{lens}$ parameter ) used by this work play a significant role in their detections \cite{roy_2024}.


We note that our analysis is an extension of the work carried out in ~\cite{marc_2025}.
We recover most of the results in ~\cite{marc_2025} for the $\Lambda$CDM model by using the CMB$+$BAO dataset: We observe that the Bayesian upper bounds relax for NH and IH compared to 1M and DH, the frequentist upper bounds are tighter than the Bayesian ones, and the minima of the PLs lie in the non-physical region for $\Lambda$CDM. 

\rthis{As noted in previous works, constraints on neutrino mass bounds are sensitive to the choice of DE parameterization. Ref.~\cite{gabriel_2025} examined the effect of the Barboza-Alcaniz (BA)~\cite{barboza_2008} and the Jassal-Bagla-Padmanabhan (JBP)~\cite{jassal_2005} DE parameterizations on neutrino mass bounds and found a correlation between the DE EOS and neutrino mass bounds. They find that the bounds on the neutrino mass are relaxed for the JBP and BA parameterizations compared to $\Lambda$CDM. Our results are consistent with these findings: we see that CPL and EXP lead to relaxed neutrino upper bounds as compared to $\Lambda$CDM.} 

\rthis{Ref.~\cite{gabriel_2025} reports that for the CMB$+$BAO$+$PantheonPlus dataset, BA and JBP parametrizaitons  yield an  upper limit of 0.18 eV and 0.13 eV, respectively. They emphasize that larger uncertainties in $w_0$ and $w_a$ values lead to tighter constraints on the neutrino mass. Further, it has been shown in \cite{giare_2024, shubham_2025} that JBP gives larger EOS uncertainties compared to BA and CPL while BA and CPL give similar constraints. This can also be seen from the evolution of $w(z)$~\cite{shubham_2025, gabriel_2025}, where CPL and BA show similar behavior but JBP differs in that it's $w(z)$ lies close to -1 after the first phantom crossing and even return to the quintessence region ($w(z) > -1$) at higher $z$. Based on these observations, it can be said that the neutrino mass constraints from CPL, BA and EXP parameterizations will be broadly similar (given their similar EOS behaviour) while JBP will show tighter upper bounds on the sum of neutrino mass.}

\rthis{The slight tightening of constraints we obtain for CPL is also in agreement with Ref.~\cite{gabriel_2025} who noted that the differences in neutrino mass bounds (in the JBP and BA parameterizations) arise from variations in the EOS constraints at higher redshifts (cf. Fig.~\ref{fig10}). The very slight relaxation in the EXP parameterization can be understood to be the effect of very similar EOS constraints; the difference arising due to the second order correction terms. A similar analysis studying the effect of neutrino mass hierarchies on the DE parameterization has also been done in \cite{yang_2017}, where it was found that neutrino mass hierarchies change DE model parameters mildly.}

To conclude, we find that even without increasing the parameter space of the DE EOS but only adding in certain terms to increase its flexibility (EXP parameterization), leads to changes in the upper bounds of the sum of neutrino masses. Further, we reaffirm the fact that the 95\% upper bounds on neutrino mass sum are  dependent on the priors used (the imposed lower limit), the DE parameterization and the dataset used. They have very mild dependence on the hierarchy, since cosmology is highly sensitive to the total neutrino mass and not on the mass splittings.
The supposed detection of positive neutrino masses is also dependent on the imposed lower limit, DE parameterization and the dataset used. With the existence of many more DE parameterizations in the literature, the 95\% upper bounds on the neutrino mass sum is dependent on the specific parameterization adopted and a more exhaustive analysis for other parametrizations  will be deferred  to a future work. With further increase in observational precision, we might be able to match limits from both terrestrial observations and cosmologically inferred neutrino mass upper bounds.

\begin{acknowledgments}
SB would like to extend his gratitude to the University Grants Commission (UGC), Govt. of India for their continuous support through the Junior Research Fellowship, which has played a crucial role in the successful completion of our research. We would also like to thank Laura Herold for assistance with the \texttt{pinc} code, Tanvi Karwal for useful correspondence, Shouvik Roy Choudhury and Eoin \'O Colg\'ain for valuable comments on this work.
Computational work was supported by the National Supercomputing Mission (NSM), Government of India, through access to the ``PARAM SEVA'' facility at IIT Hyderabad. The NSM is implemented by the Centre for Development of Advanced Computing (C-DAC) with funding from the Ministry of Electronics and Information Technology (MeitY) and the Department of Science and Technology (DST). We also acknowledge the use of IUCAA HPC Computing facilities (project ID - hpc2502002). We also acknowledge the anonymous referee for useful feedback and comments on the manuscript.
\end{acknowledgments}

\clearpage 
\appendix
\section{Results for the  other cosmological parameters}
\label{appendixA}

We now make a few generic observations from the above analyses on the  other cosmological parameters (in addition to the neutrino mass).
We find no significant variations in the values of $\ln(10^{10}A_{s })$, $n_s$,  and $\tau_\mathrm{reio}$, irrespective of the model, dataset and hierarchy combinations, as can be seen from Tables~\ref{table6}, \ref{table7}, \ref{table8} and \ref{table9}.

For a particular hierarchy, there is a slight tension among the $H_0$ values depending on the model and dataset used. For all cases, CPL and EXP models give consistent  $H_0$ values within 1$\sigma$.  However, there is a $\gtrsim2\sigma$ tension when considering the Hubble constant values between $\Lambda$CDM and CPL/EXP in the case of CMB$+$BAO dataset across all hierarchies. This reduces to $\gtrsim 1\sigma$ when we include SNe dataset. Including the SNe dataset shifts $H_0$ to higher values by $(1-2)\sigma$ in the case of CPL/EXP models in all hierarchy cases. This does not happen for the $\Lambda$CDM case.

When considering CMB$+$BAO dataset (for all hierarchies), $\Omega_m$ values increase by $\gtrsim2 \sigma$ in CPL/EXP parameterizations compared to $\Lambda$CDM. Inclusion of SNe dataset reconcile $\Omega_m$ values to within $1\sigma$ for all the three parameterizations. $\Omega_m$ for $\Lambda$CDM remains consistent (with the CMB$+$BAO dataset case) while for CPL and EXP they decrease when including SNe. 

We also note that for a particular model and dataset, $w_0$ and $w_a$ values are consistent within $1\sigma$ for different neutrino mass hierarchies. This is also true for $\sigma_8$ parameter, which is anti-correlated with $M_\mathrm{tot}$.

\begin{table}[htbp!]
\caption{Cosmological parameter constraints at 68\% credible intervals for 1M neutrino mass hierarchy.}
\label{table6}
\centering
    \begin{tabular}{c@{\hspace{0.2cm}}c@{\hspace{0.2cm}}c@{\hspace{0.2cm}}c@{\hspace{0.2cm}}c@{\hspace{0.2cm}}c@{\hspace{0.2cm}}c@{\hspace{0.2cm}}c@{\hspace{0.2cm}}c}
    \hline
    \thead{Model/Dataset} & \thead{$H_0$} & \thead{$\Omega_m$} & \thead{$\omega_0$} & \thead{$\omega_a$} & \thead{$\ln10^{10}A_{s }$} & \thead{$n_s$} & \thead{$\tau_\mathrm{reio}$}\\
    & (km/sec/Mpc)& & & & & & \\
    \hline
    \hline
    & & & & & & & \\[0.5ex]  
    \textbf{CMB$+$BAO} & & & & & & & \\[1.5ex]
    $\Lambda$CDM & $68.61_{-0.31}^{+0.32}$ & $0.2992\pm0.0039$           & ---                     & ---                     & $3.050_{-0.016}^{+0.014}$ & $0.9704_{-0.0035}^{+0.0034}$ & $0.0587_{-0.0079}^{+0.0071}$ \\[1ex]
    CPL          & $63.5\pm2.3$         & $0.3562_{-0.03}^{+0.024}$ & $-0.38_{-0.3}^{+0.24}$ & $-1.90_{-0.68}^{+0.98}$ & $3.043\pm0.015$           & $0.9664_{-0.0039}^{+0.0038}$ & $0.0539_{-0.008}^{+0.0074}$ \\[1ex]
    EXP          & $63.9_{-2.1}^{+2.3}$ & $0.352_{-0.028}^{+0.022}$ & $-0.5_{-0.28}^{+0.2}$ & $-1.40_{-0.49}^{+0.74}$ & $3.042\pm0.015$          & $0.9664\pm0.0038$           & $0.0537_{-0.0079}^{+0.0074}$\\[1ex]
    
    \textbf{CMB$+$BAO$+$PantheonPlus} & & & & & & & \\[1.5ex]
    $\Lambda$CDM & $68.5_{-0.3}^{+0.32}$ & $0.3008_{-0.0039}^{+0.0037}$ & ---                     & ---                     & $3.049_{-0.016}^{+0.014}$ & $0.9698_{-0.0033}^{+0.0036}$ & $0.0582_{-0.0079}^{+0.0068}$ \\[1ex]
    CPL          & $67.59_{-0.64}^{+0.61}$ & $0.311_{-0.0061}^{+0.006}$ & $-0.840_{-0.058}^{+0.054}$ & $-0.6_{-0.2}^{+0.25}$ & $3.045_{-0.015}^{+0.014}$ & $0.9679_{-0.0037}^{+0.0037}$            & $0.0552_{-0.0078}^{+0.0072}$ \\[1ex]
    EXP          & $67.64\pm0.62$          & $0.3107_{-0.0062}^{+0.0058}$ & $-0.850_{-0.054}^{+0.053}$          & $-0.46_{-0.16}^{+0.19}$ & $3.044\pm0.015$ & $0.9677_{-0.0037}^{+0.0038}$ & $0.055_{-0.0079}^{+0.0071}$ \\[1ex]
    
    \textbf{CMB$+$BAO$+$DESY5} & & & & & & & \\[1.5ex]
    $\Lambda$CDM & $68.38\pm0.31$          & $0.3022\pm0.0038$ & ---                     & ---                     & $3.048_{-0.015}^{+0.014}$ & $0.9694_{-0.0035}^{+0.0034}$ & $0.0577_{-0.0078}^{+0.0069}$ \\[1ex]
    CPL          & $66.89_{-0.59}^{+0.57}$ & $0.3181_{-0.0059}^{+0.006}$ & $-0.758_{-0.061}^{+0.058}$ & $-0.82_{-0.24}^{+0.27}$ & $3.044\pm0.015$ & $0.9674\pm0.0038$ & $0.0547_{-0.0079}^{+0.0072}$ \\[1ex]
    EXP          & $66.91_{-0.59}^{+0.56}$ & $0.3181_{-0.006}^{+0.0058}$ & $-0.775_{-0.057}^{+0.054}$ & $-0.64_{-0.17}^{+0.23}$ & $3.044\pm0.015$ & $0.9673_{-0.0037}^{+0.0038}$ & $0.0547_{-0.0077}^{+0.0074}$ \\[1ex]
    & & & & & & & \\[0.5ex]  
    \hline
    \end{tabular}
\end{table}

\begin{table}[htbp!]
\caption{Cosmological parameter constraints at 68\% credible intervals for degenerate neutrino mass hierarchy.}
\label{table7}
\centering
    \begin{tabular}{c@{\hspace{0.2cm}}c@{\hspace{0.2cm}}c@{\hspace{0.2cm}}c@{\hspace{0.2cm}}c@{\hspace{0.2cm}}c@{\hspace{0.2cm}}c@{\hspace{0.2cm}}c@{\hspace{0.2cm}}c}
    \hline
    \thead{Model/Dataset} & \thead{$H_0$} & \thead{$\Omega_m$} & \thead{$\omega_0$} & \thead{$\omega_a$} & \thead{$\ln10^{10}A_{s }$} & \thead{$n_s$} & \thead{$\tau_\mathrm{reio}$}\\
    & (km/sec/Mpc)& & & & & & \\
    \hline
    \hline
    & & & & & & & \\[0.5ex]  
    \textbf{CMB$+$BAO} & & & & & & & \\[1.5ex]
    $\Lambda$CDM & $68.6_{-0.3}^{+0.33}$  & $0.2996_{-0.004}^{+0.0038}$ & --- & --- & $3.049_{-0.015}^{+0.014}$ & $0.9694_{-0.0033}^{+0.0035}$ & $0.058_{-0.0075}^{+0.007}$ \\[1ex]
    CPL          & $63.6_{-2.3}^{+2.2}$  & $0.355_{-0.029}^{+0.024}$ & $-0.39_{-0.29}^{+0.23}$ & $-1.86_{-0.67}^{+0.93}$ & $3.042_{-0.015}^{+0.014}$ & $0.9656_{-0.0036}^{+0.0038}$ & $0.0535_{-0.0077}^{+0.0074}$ \\[1ex]
    EXP          & $63.9_{-2.2}^{+2.2}$  & $0.351_{-0.029}^{+0.022}$ & $-0.46_{-0.27}^{+0.21}$ & $-1.39_{-0.49}^{+0.75}$ & $3.041_{-0.015}^{+0.014}$ & $0.9654\pm0.0038$ & $0.0532_{-0.0077}^{+0.0072}$ \\[1ex]
    
    \textbf{CMB$+$BAO$+$PantheonPlus} & & & & & & & \\[1.5ex]
    $\Lambda$CDM & $68.45\pm0.31$  & $0.3011\pm0.0038$ & --- & --- & $3.048_{-0.015}^{+0.014}$ & $0.9688_{-0.0034}^{+0.0033}$ & $0.0577_{-0.0074}^{+0.0069}$ \\[1ex]
    CPL          & $67.57_{-0.62}^{+0.59}$  & $0.3111_{-0.006}^{+0.0058}$ & $-0.841_{-0.057}^{+0.056}$ & $-0.57_{-0.21}^{+0.24}$ & $3.043_{-0.015}^{+0.014}$ & $0.9668_{-0.0038}^{+0.0036}$ & $0.0547_{-0.0076}^{+0.0072}$  \\[1ex]
    EXP          & $67.6_{-0.61}^{+0.59}$   & $0.311_{-0.006}^{+0.0057}$ & $-0.850_{-0.055}^{+0.049}$ & $-0.46_{-0.16}^{+0.19}$ & $3.043_{-0.015}^{+0.014}$ & $0.9667\pm0.0037$ & $0.0547_{-0.0079}^{+0.0072}$ \\[1ex]
    
    \textbf{CMB$+$BAO$+$DESY5} & & & & & & & \\[1.5ex]
    $\Lambda$CDM & $68.35_{-0.31}^{+0.32}$ & $0.3024_{-0.0039}^{+0.0038}$ & --- & --- & $3.048_{-0.015}^{+0.013}$ & $0.9684_{-0.0033}^{+0.0034}$ & $0.0574_{-0.0074}^{+0.0067}$ \\[1ex]
    CPL          & $66.87_{-0.57}^{+0.58}$ & $0.3183_{-0.0062}^{+0.0055}$ & $-0.76\pm0.06$ & $-0.82_{-0.23}^{+0.27}$ & $3.042_{-0.015}^{+0.014}$ & $0.9663_{-0.0038}^{+0.0037}$ & $0.0541_{-0.0078}^{+0.0072}$ \\[1ex]
    EXP          & $66.89_{-0.58}^{+0.57}$ & $0.3182_{-0.0062}^{+0.0058}$ & $-0.778_{-0.06}^{+0.052}$ & $-0.63_{-0.17}^{+0.22}$ & $3.042\pm0.015$ & $0.9663_{-0.0039}^{+0.0037}$ & $0.0539_{-0.0078}^{+0.0074}$ \\[1ex]
    & & & & & & & \\[0.5ex]  
    \hline
    \end{tabular}
\end{table}

\begin{table}[htbp!]
\caption{Cosmological parameter constraints at 68\% credible intervals for the normal neutrino mass hierarchy.}
\label{table8}
\centering
    \begin{tabular}{c@{\hspace{0.2cm}}c@{\hspace{0.2cm}}c@{\hspace{0.2cm}}c@{\hspace{0.2cm}}c@{\hspace{0.2cm}}c@{\hspace{0.2cm}}c@{\hspace{0.2cm}}c@{\hspace{0.2cm}}c}
    \hline
    \thead{Model/Dataset} & \thead{$H_0$} & \thead{$\Omega_m$} & \thead{$\omega_0$} & \thead{$\omega_a$} & \thead{$\ln10^{10}A_{s }$} & \thead{$n_s$} & \thead{$\tau_\mathrm{reio}$}\\
    & (km/sec/Mpc)& & & & & & \\
    \hline
    \hline
    & & & & & & & \\[0.5ex]  
    \textbf{CMB$+$BAO} & & & & & & & \\[1.5ex]
    $\Lambda$CDM & $68.34_{-0.32}^{+0.31}$ & $0.3017_{-0.004}^{+0.0039}$ & --- & --- & $3.055_{-0.017}^{+0.015}$ & $0.9706\pm0.0035$ & $0.0618_{-0.0086}^{+0.0074}$ \\[1ex]
    CPL          & $63.0_{-2.2}^{+2.4}$ & $0.363_{-0.031}^{+0.023}$ & $-0.31_{-0.31}^{+0.24}$ & $-2.14_{-0.68}^{+0.98}$ & $3.043_{-0.016}^{+0.015}$ & $0.9654_{-0.0038}^{+0.0037}$ & $0.0544_{-0.0079}^{+0.0073}$ \\[1ex]
    EXP          & $63.5_{-2.3}^{+1.9}$ & $0.357_{-0.025}^{+0.024}$ & $-0.41\pm0.23$ & $-1.54_{-0.56}^{+0.61}$ & $3.043_{-0.015}^{+0.014}$ & $0.9654\pm0.0037$ & $0.0543_{-0.0077}^{+0.0072}$\\[1ex]
    
    \textbf{CMB$+$BAO$+$PantheonPlus} & & & & & & & \\[1.5ex]
    $\Lambda$CDM & $68.3_{-0.3}^{+0.31}$ & $0.3029_{-0.0039}^{+0.0038}$ & --- & --- & $3.054\pm0.015$ & $0.9701_{-0.0035}^{+0.0033}$ & $0.0611_{-0.008}^{+0.0071}$ \\[1ex]
    CPL          & $67.6_{-0.64}^{+0.6}$ & $0.312_{-0.006}^{+0.0059}$ & $-0.829_{-0.056}^{+0.057}$ & $-0.66_{-0.21}^{+0.24}$ & $3.046_{-0.016}^{+0.014}$ & $0.9671\pm0.0037$ & $0.056_{-0.0083}^{+0.007}$ \\[1ex]
    EXP          & $67.56_{-0.61}^{+0.59}$ & $0.3121_{-0.0059}^{+0.0058}$ & $-0.841_{-0.053}^{+0.052}$ & $-0.52_{-0.16}^{+0.19}$ & $3.046_{-0.016}^{+0.014}$ & $0.967_{-0.0038}^{+0.0036}$ & $0.0560_{-0.0079}^{+0.0067}$ \\[1ex]
    
    \textbf{CMB$+$BAO$+$DESY5} & & & & & & & \\[1.5ex]
    $\Lambda$CDM & $68.14\pm0.29$ & $0.3044\pm0.0037$ & --- & --- & $3.053_{-0.016}^{+0.014}$ & $0.9695\pm0.0034$ & $0.0603_{-0.008}^{+0.0071}$ \\[1ex]
    CPL          & $66.82_{-0.61}^{+0.55}$ & $0.3195_{-0.0058}^{+0.0059}$ & $-0.746_{-0.063}^{+0.056}$ & $-0.91_{-0.22}^{+0.27}$ & $3.045_{-0.016}^{+0.014}$ & $0.9666_{-0.0038}^{+0.0037}$ & $0.0557_{-0.0079}^{+0.0071}$ \\[1ex]
    EXP          & $66.85_{-0.57}^{+0.56}$ & $0.3193_{-0.0059}^{+0.0056}$ & $-0.764_{-0.058}^{+0.053}$ & $-0.72_{-0.17}^{+0.22}$ & $3.045_{-0.015}^{+0.014}$ & $0.9666_{-0.0036}^{+0.0037}$ & $0.0555_{-0.0079}^{+0.0072}$ \\[1ex]
    & & & & & & & \\[0.5ex]  
    \hline
    \end{tabular}
\end{table}

\begin{table}[htbp!]
\caption{Cosmological parameter constraints at 68\% credible intervals for inverted neutrino mass hierarchy.}
\label{table9}
\centering
    \begin{tabular}{c@{\hspace{0.2cm}}c@{\hspace{0.2cm}}c@{\hspace{0.2cm}}c@{\hspace{0.2cm}}c@{\hspace{0.2cm}}c@{\hspace{0.2cm}}c@{\hspace{0.2cm}}c@{\hspace{0.2cm}}c}
    \hline
    \thead{Model/Dataset} & \thead{$H_0$} & \thead{$\Omega_m$} & \thead{$\omega_0$} & \thead{$\omega_a$} & \thead{$\ln10^{10}A_{s }$} & \thead{$n_s$} & \thead{$\tau_\mathrm{reio}$}\\
    & (km/sec/Mpc)& & & & & & \\
    \hline
    \hline
    & & & & & & & \\[0.5ex]  
    \textbf{CMB$+$BAO} & & & & & & & \\[1.5ex]
    $\Lambda$CDM & $68.2_{-0.3}^{+0.31}$ & $0.3035_{-0.004}^{+0.0038}$ & --- & --- & $3.06_{-0.017}^{+0.015}$ & $0.9717_{-0.0034}^{+0.0033}$ & $0.0646_{-0.0087}^{+0.0072}$\\[1ex]
    CPL          & $62.5_{-2.3}^{+2.3}$ & $0.369_{-0.031}^{+0.024}$ & $-0.24_{-0.31}^{+0.24}$ & $-2.38_{-0.71}^{+0.94}$ & $3.046_{-0.015}^{+0.014}$ & $0.9656\pm0.0037$ & $0.0554_{-0.0078}^{+0.0071}$ \\[1ex]
    EXP          & $62.9_{-2.1}^{+2.2}$ & $0.364_{-0.028}^{+0.023}$ & $-0.34_{-0.28}^{+0.22}$ & $-1.76_{-0.51}^{+0.75}$ & $3.045_{-0.016}^{+0.014}$ & $0.9657_{-0.0036}^{+0.0039}$ & $0.055_{-0.008}^{+0.007}$ \\[1ex]
    
    \textbf{CMB$+$BAO$+$PantheonPlus} & & & & & & & \\[1.5ex]
    $\Lambda$CDM & $68.06_{-0.3}^{+0.29}$ & $0.3047_{-0.0038}^{+0.0037}$ & --- & --- & $3.059_{-0.018}^{+0.014}$ & $0.9713_{-0.0037}^{+0.0033}$ & $0.0641_{-0.0089}^{+0.007}$ \\[1ex]
    CPL          & $67.51_{-0.61}^{+0.62}$ & $0.313\pm0.006$ & $-0.818_{-0.057}^{+0.058}$ & $-0.75_{-0.22}^{+0.23}$ & $3.049_{-0.016}^{+0.014}$ & $0.9674\pm0.0037$ & $0.0579_{-0.008}^{+0.0071}$ \\[1ex]
    EXP          & $67.5\pm0.6$ & $0.314\pm0.006$ & $-0.829_{-0.054}^{+0.053}$ & $-0.6_{-0.16}^{+0.19}$ & $3.049_{-0.016}^{+0.014}$ & $0.9672_{-0.0037}^{+0.0036}$ & $0.058_{-0.0081}^{+0.007}$ \\[1ex]
    
    \textbf{CMB$+$BAO$+$DESY5} & & & & & & & \\[1.5ex]
    $\Lambda$CDM & $67.96_{-0.3}^{+0.29}$ & $0.3061_{-0.0037}^{+0.0038}$ & --- & --- & $3.058_{-0.018}^{+0.014}$ & $0.9708_{-0.0035}^{+0.0033}$ & $0.0634_{-0.0089}^{+0.0068}$ \\[1ex]
    CPL          & $66.75\pm0.57$ & $0.3207_{-0.0058}^{+0.0059}$ & $-0.732_{-0.062}^{+0.059}$ & $-1.00_{-0.23}^{+0.27}$ & $3.048_{-0.015}^{+0.014}$ & $0.9671_{-0.0036}^{+0.0037}$ & $0.0572_{-0.0081}^{+0.0071}$ \\[1ex]
    EXP          & $66.81_{-0.57}^{+0.56}$ & $0.3203_{-0.0059}^{+0.0056}$ & $-0.756_{-0.058}^{+0.054}$ & $-0.78_{-0.19}^{+0.21}$ & $3.048_{-0.015}^{+0.014}$ & $0.9668_{-0.0037}^{+0.0035}$ & $0.0570_{-0.0079}^{+0.0071}$ \\[1ex]
    & & & & & & & \\[0.5ex]  
    \hline
    \end{tabular}
\end{table}

\clearpage

\bibliography{references}

\end{document}